\documentclass[aip,jcp,reprint,noshowkeys]{revtex4-2}
\usepackage{graphicx,float,dcolumn,bm,xcolor,microtype,multirow,amscd,amsmath,amssymb,amsfonts,physics,upgreek,pifont,wrapfig,enumitem,siunitx}
\usepackage[justification=raggedright,singlelinecheck=false]{caption} 
\usepackage[version=4]{mhchem}
\usepackage[utf8]{inputenc}
\usepackage[T1]{fontenc}
\usepackage{braket} 
\usepackage[sep=0em, offset=1.6em]{simpler-wick}  
\usepackage{natbib}

\newcommand{\inlinecite}[1]{\citenum{#1}}

\usepackage{hyperref}
\hypersetup{
    colorlinks=true,
    linkcolor=blue,
    filecolor=blue,      
    urlcolor=blue,	
    citecolor=blue
}

\usepackage{caption} 
\usepackage{subcaption}
\usepackage{tabularx}
\usepackage{cleveref}

\usepackage{float} 
\newcommand{\UOX}{Physical and Theoretical Chemical Laboratory, University of Oxford, South Parks Road, Oxford, OX1 3QZ, U.K.}

\begin{document}

\title{Spin-free Generalised Normal Ordered Coupled Cluster}
\author{Nicholas~Lee}
\email{nicholas.lee@chem.ox.ac.uk}
\affiliation{\UOX}

\author{David~P.~Tew}
\email{david.tew@chem.ox.ac.uk}
\affiliation{\UOX}

\date{\today}

\begin{abstract}
We present a spin-free, size-extensive, and size-consistent coupled cluster method based on a generalised normal ordered exponential ansatz. This approach is a natural generalisation of single-reference coupled cluster theory for arbitrary spin eigenfunctions. The working equations are size-extensive through the generalised normal order formalism, and made spin-free with the spin-ensemble approach. Redundancies amongst excitations are eliminated by selecting only those excitations that project the reference function onto the first-order interacting space. Furthermore, by utilising localised orbitals, the proposed method describes dissociation into open-shell fragments size-consistently. Numerical results on prototypical multireference systems at the singles and doubles level of theory are competitive with existing multireference approaches, yet with more compact working equations.
\end{abstract}

\maketitle
\raggedbottom

\section{Introduction}
Application of coupled cluster theory to (quasi-)degenerate electronic states where multiple determinants are necessary remains a challenge for electronic structure theory\cite{lyakhMultireferenceNatureChemistry2012,carskyRecentProgressCoupled,kohnStatespecificMultireferenceCoupledcluster2013,evangelistaPerspectiveMultireferenceCoupled2018}. Many early attempts at multireference coupled cluster (MRCC) involve solving for the ground and excited states using an effective Hamiltonian in a model space \cite{shavittManyBodyMethodsChemistry2009}. These attempts are typically classified either as state-universal\cite{piecuchStateUniversalMultiReferenceCoupledCluster2002,paldusAnalysisMultireferenceStateuniversal2003} (SU-MRCC) when the determinants making up the model space have the same number of electrons, or valence-universal\cite{haqueApplicationClusterExpansion1984,stolarczykCoupledclusterMethodFock1985,stolarczykCoupledclusterMethodFock1985a,stolarczykCoupledclusterMethodFock1988,stolarczykCoupledclusterMethodFock1988a,lindgrenNoteLinkedDiagramCoupledCluster1985,lindgrenConnectivityCriteriaOpenshell1987,kaldorFockSpaceCoupled1991} (VU-MRCC) when the determinants can have variable number of electrons. These effective Hamiltonian approaches are usually beset by intruder state problems\cite{hoseDiagrammaticManybodyPerturbation1979,hoseGeneralModelSpaceDiagrammaticPerturbation1980,hoseQuasidegeneratePerturbationTheory1982,kaldorIntruderStatesIncomplete1988} when model space determinants are strongly coupled to those in the complementary space. The problem of intruder states has led to the development of intermediate Hamiltonian approaches\cite{malrieuIntermediateHamiltoniansNew1985}, where a subset of the effective Hamiltonian eigenstates are targeted. 
When solving only for one eigenstate, the intermediate Hamiltonian approach is known as state-specific. State-specific approaches in MRCC have been extensively developed over the last few decades\cite{mahapatraStateSpecificMultiReferenceCoupled1998,kohnStatespecificMultireferenceCoupledcluster2013}. In general, these approaches can be broadly divided into two classes (i) Methods which use a wave operator for each determinant in the reference function (Jeziorski-Monkhorst (JM) ansatz\cite{jeziorskiCoupledclusterMethodMultideterminantal1981}) and (ii) A single wave operator is applied onto a given reference function as pioneered by Mukherjee\cite{mukherjeeNonperturbativeOpenshellTheory1975}. Common to these approaches is that there are usually more excitation operators than there exist excited states\cite{kohnStatespecificMultireferenceCoupledcluster2013,evangelistaPerspectiveMultireferenceCoupled2018}. This leads to a redundancy problem where the number of residual equations, found by projecting the Schrödinger equation onto excited states, is fewer than the number of excitation operators. The corresponding cluster amplitudes are therefore underdetermined. With (i), the JM ansatz, it is typical to employ sufficiency conditions to overcome such redundancies. The different choices of sufficiency conditions employed\cite{kongConnectionFewJeziorskiMonkhorst2009,lyakhMultireferenceNatureChemistry2012} define the different Hilbert-Space MRCC (HS-MRCC) methods.\cite{hanrathExponentialMultireferenceWavefunction2005,hanrathExponentialMultireferenceWavefunction2008,hubacUseBrillouinWignerPerturbation2000,mahapatraStateSpecificMultiReferenceCoupled1998,mahapatraSizeconsistentStatespecificMultireference1999,pittnerContinuousTransitionBrillouin2003,evangelistaHighorderExcitationsStateuniversal2006,sinhaDevelopmentApplicationsUnitary2012,senFormulationImplementationUnitary2012,chakravartiSystematicImprovementUGASSMRCCSD2023}However, many of these methods do not correspond to solving a projected Schrödinger equation (the proper residual equation condition is not satisfied\cite{kongConnectionFewJeziorskiMonkhorst2009}). With (ii), initial attempts by Mukherjee\cite{mukherjeeNonperturbativeOpenshellTheory1975,mukherjeeCorrelationProblemOpenshell1975,mukherjeeAbinitioDerivationPielectron1977,mukherjeeApplicationsNonperturbativeManybody1977} and coworkers, as well as Banerjee and Simons\cite{banerjeeCoupledclusterMethodMulticonfiguration1981} did not explicitly consider the problem of redundancy. A method of handling this redundancy was introduced by Evangelista and Gauss\cite{evangelistaOrbitalinvariantInternallyContracted2011} involves constructing an overlap metric and removing the null space to identify linearly independent excitation operators. This is known as internally contracted MRCC (ic-MRCC). A correctly scaling implementation of this method and the development of several low-cost approximations were further advanced by Köhn and co-workers\cite{hanauerPilotApplicationsInternally2011,hanauerPerturbativeTreatmentTriple2012,hanauerCommunicationRestoringFull2012,evangelistaSequentialTransformationApproach2012,samantaExcitedStatesInternally2014,aotoInternallyContractedMultireference2016,samantaFirstorderPropertiesInternally2018,kohnSecondorderApproximateInternally2019,blackLinearQuadraticInternally2019,kohnImprovedSimplifiedOrthogonalisation2020,blackEfficientImplementationInternally2023,zielinskiPerformanceTestsSecondOrder2023,waigumAccurateThermochemistryMultireference2024,adamMultireferenceCoupledClusterTheory2025}. Since these redundancy problems arise from trying to solve a series of projective equations, another approach to overcome this difficulty is through constructing a different set of equations. The working equations are found by expanding a similarity transformed Hamiltonian in terms of  $n$-body operators and requiring that those corresponding to the excitations in the cluster operator vanish.\cite{stolarczykCompleteActiveSpace1994,nooijenGeneralSpinAdaptation1996}. Formally, this solves the problem of linear dependencies as there are as many residuals as there are cluster amplitudes. This approach is known as the many-body residual method and has been adopted in several formalisms \cite{stolarczykCompleteActiveSpace1994,nooijenGeneralMultireferenceCoupled2001,dattaStatespecificPartiallyInternally2011,nooijenCommunicationMultireferenceEquation2014,feldmannRenormalizedInternallyContracted2024}, including partially internally contracted MRCC (pic-MRCC) and multireference equation-of-motion coupled-cluster (MREOM-CC).

The aforementioned MRCC methods are known as \emph{genuine} MRCC methods because they allow for relaxation of the reference function parameters. A different way of circumventing the redundancy problem is to use a fixed reference function, and therefore a fixed set of linearly independent excitation operators. These are known as \emph{alt}-MRCC methods\cite{oliphantMultireferenceCoupledCluster1993,oliphantMultireferenceCoupledCluster1993,piecuchStateselectiveMultireferenceCoupledcluster1993,lyakhGeneralizationStatespecificCompleteactivespace2008,kallayGeneralStateselectiveMultireference2002}. This was pioneered by Oliphant and Adamowicz, who used a complete active space (CAS) reference with single and double excitations from each of the reference determinants\cite{oliphantMultireferenceCoupledCluster1993,oliphantMultireferenceCoupledCluster1993}. 
Similar to this are the Nakatsuji's multireference symmetry-adapted cluster\cite{nakatsujiMultireferenceClusterExpansion1985} (MR-SAC), and Li and Paldus' unitary group approach coupled cluster\cite{liMulticonfigurationalSpinadaptedSinglereference1993,liAutomationImplementationSpinadapted1994,liUnitaryGroupBased1997,liUnitarygroupbasedOpenshellCoupledcluster1998} (UGA-CC), which use multi-configurational references with a fully spin-adapted treatment. While spin-adaptation was performed with spin-projection operators in MR-SAC, spin-free operators were constructed through a unitary group approach in UGA-CC. More recently, the UGA-CC approach has been generalised to arbitrary high-spin open-shell systems by Herrmann and Hanrath\cite{herrmannGenerationSpinadaptedSpincomplete2020,herrmannCorrectlyScalingRigorously2022,herrmannAnalysisDifferentSets2022}. Other wavefunction ansätze have also been utilised for multireference coupled cluster methods, including matrix product states (Block-correlated coupled cluster)\cite{liBlockcorrelatedCoupledCluster2004,wangDescribingStrongCorrelation2020} and generalised valence bond functions (Multireference ring coupled cluster)\cite{szabadosRingCoupledclusterDoubles2017,margocsyRingCoupledCluster2020}. \\

Another challenging issue with MRCC lies in the complexity of the working equations. This is due to the presence of excitation operators, which can excite from and into active orbitals (orbitals which are partially occupied in the reference state), and hence do not commute. This in turn gives rise to high commutator ranks in the Baker-Campbell-Hausdorff (BCH) expansion\cite{helgakerMolecularElectronicStructureTheory} of the coupled cluster equations due to contractions between excitation operators $\hat{T}$. For instance, the inclusion of singles and doubles can lead to eight-fold commutators\cite{hanauerPilotApplicationsInternally2011}. This is prohibitively expensive for all but the smallest systems, and it is typical to truncate the equations at a lower commutator order. In ic-MRCC, the equations are usually truncated at the two-fold commutator\cite{evangelistaOrbitalinvariantInternallyContracted2011} and have been shown to deliver promising numerical results with reasonable computational costs.\\

Other efforts to reduce equation complexity aim to address the non-commutativity of the excitation operator. To that end, Lindgren proposed a normal-ordered exponential (NOE) wave operator\cite{lindgrenCoupledclusterApproachManybody1978,lindgrenLinkedDiagramCoupledClusterExpansions1985,lindgrenNoteLinkedDiagramCoupledCluster1985}. As a consequence of Wick's theorem\cite{wickEvaluationCollisionMatrix1950,shavittManyBodyMethodsChemistry2009}, no contractions between the excitation operators are possible. The NOE ansatz, therefore, benefits from simpler working equations. This ansatz has found applications in VU-MRCC\cite{kaldorFockSpaceCoupled1991,kaldorIntruderStatesIncomplete1988} and Similarity Transformed Equation-of-Motion Coupled Cluster (STEOM-CC)\cite{nooijenManybodySimilarityTransformations1996,nooijenSimilarityTransformedEquationofmotion1997,nooijenNewMethodExcited1997,nooijenSimilarityTransformedEquation1999} of Nooijen and Bartlett by decoupling excitations belonging to different valence sectors. This has later been adopted by Mukherjee and co-workers for state-specific applications. This led to the development of two methods, the combinatoric open-shell coupled cluster (COS-CC)\cite{janaDevelopmentApplicationsRelaxationinducing1999,janaCompactSpinfreeCluster2002,janaUseNewCluster2002} and the unitary group approach open-shell coupled cluster (UGA-OSCC)\cite{senInclusionOrbitalRelaxation2018,chakravartiReappraisalNormalOrdered2021}. These two methods differ in their treatment of the overparameterisation of the cluster amplitudes. COS-CC employs the redefinition of excitation operators such that distinct excitations appear uniquely, and only allows for certain classes of excitation operators to contract amongst themselves. Meanwhile, UGA-OSCC uses a predetermined set of linearly independent excitation operators. UGA-OSCC thus generalises a previous effort\cite{nooijenGeneralMultireferenceCoupled2001} by Nooijen and Lotrich to correlate open-shell doublet states with a NOE ansatz. More recently, the authors have also used a NOE to correlate a single CSF\cite{gunasekeraMultireferenceCoupledCluster2024} in a manner similar to the UGA-OSCC method, albeit with different approximations to the working equations. \\

Within the NOE framework, one has the freedom to choose a reference with which a normal ordering is defined. In principle, the complete coupled cluster equations will be agnostic to the chosen reference. In practice, however, the equations have to be truncated to be computationally tractable and the truncated equations may differ for different normal orderings.
Two common choices of reference are (i) the closed-shell vacuum and (ii) the given reference function. The former is routinely used in closed-shell coupled cluster codes, and the latter choice is known as Generalised Normal Ordering\cite{kutzelniggNormalOrderExtended1997} (GNO, alternatively known as KM normal ordering or extended normal ordering in various literature). This particular definition of normal ordering has been used in several modern electronic structure methods\cite{yanaiCanonicalTransformationTheory2007,liSpinfreeFormulationMultireference2021,margocsyRingCoupledCluster2020,feldmannRenormalizedInternallyContracted2024}, such as canonical transformation theory of Yanai and Chan, and the driven similarity renormalisation group (DSRG) method of Li and Evangelista. A benefit of this ordering is that the equations under GNO are connected, which is a prerequisite for size-extensivity\cite{mukherjeeCorrelationProblemOpenshell1975,lindgrenConnectivityCriteriaOpenshell1987}. 

In this article, we present a spin-free coupled cluster method based on the NOE ansatz, with the exponential operator normal ordered with respect to the given reference (GNO). Accordingly, we shall refer to our proposed method as Generalised Normal Ordered Coupled Cluster (GNOCC). We shall focus on using CASSCF wavefunctions as the reference. However, the method admits any spin eigenfunctions as a reference state. In the spirit of alt-MRCC methods, we focus on a single reference function composed of multiple determinants, where the coefficients for each determinant are kept fixed rather than re-optimised during each coupled cluster iteration.

The paper is organised as follows: In Section II, we first review the ideas behind GNO and spin-adaptation using a spin-ensemble approach. This is followed by an exposition of the GNOCC formalism, and the method through which linear dependencies in the excitation basis are handled is described. We conclude the theory section with a discussion of the properties of our prescribed theoretical framework. In Section III, we report the results with the helium atom and some high-spin open-shell systems, which serve as sanity checks for our method. We proceed to report on results for several prototypical multireference systems, highlighting performance for singlet-triplet gaps,  bond dissociation curves, and size-consistency.

\section{Theory}
We begin by considering a multi-determinant spin eigenfunction $\ket{\Phi}$. The orbitals characterising its constituent determinants can be partitioned into three classes: Core (doubly occupied in all the determinants, labelled $\mathbb{C}$), Virtual (unoccupied in all the determinants, labelled $\mathbb{V}$), and Active (orbitals which do not fall in either of the previous categories, labelled $\mathbb{A}$).\\
We shall use the indices $i,j,k,l,...$ to denote core orbitals, $a,b,c,d,...$ for virtual orbitals, $t,u,v,w,...$ for active orbitals, and $p,q,r,s,...$ for general orbitals. These orbital labels correspond to \emph{spin free} orbitals. Spin orbitals are labelled $p_{\mu}$ where Greek letters $\mu, \nu, \sigma, \tau, ...$ denote $\pm \frac{1}{2}$ electron spin functions.

\subsection{Generalised Normal Ordering}
The concept of normal ordering has been generalised by Kutzelnigg and Mukherjee for an arbitrary reference state\cite{kutzelniggNormalOrderExtended1997}, and is now known as \emph{generalised normal ordering} (GNO). We shall recapitulate the salient points of GNO here. 
Given elementary operators $\hat{A}, \hat{B}, \hat{C}, \cdots$ which represent either creation or annihilation operators, the operator product \emph{normal ordered with respect to the reference} $\ket{\Phi}$, $\{ \hat{A} \hat{B} \hat{C} \cdots \}$ satisfies the property
\begin{equation}
\braket{ \Phi | \{ \hat{A} \hat{B} \hat{C} \cdots \} | \Phi } = 0
\end{equation}
The braces $\{ \cdots \} $ denote normal ordering of the operator product. An operator product that satisfies the above condition is said to be in GNO. To place a one-body excitation operator $\hat{a}^{\dagger}_{p_{\mu}} \hat{a}_{q_{\nu}}$ in GNO, one applies Wick's theorem\cite{wickEvaluationCollisionMatrix1950,kongAlgebraicProofGeneralized2010}, 
\begin{equation}
\hat{a}^{p_{\mu} }_{q_{\nu} }		=		\{ \hat{a}^{p_{\mu} }_{q_{\nu} }		 \} + \wick{\c {\hat{a}_{p_{\mu}}^{\dagger} } \c { \hat{a}_{q_{\nu}} } } 
\end{equation}
We use the notation 
\begin{equation}
\hat{a}^{p_{\mu} q_{\nu} \cdots}_{r_{\sigma} s_{\tau} \cdots} = \hat{a}^{\dagger}_{p_{\mu}} \hat{a}^{\dagger}_{q_{\nu}} 	\cdots \hat{a}_{s_{\tau}}	
\hat{a}_{r_{\sigma}}
\end{equation}
and $\wick{\c {\hat{a}_{p_{\mu}}^{\dagger} } \c { \hat{a}_{q_{\nu}} } } $ represents a contraction between the elementary operators $\hat{a}_{p_{\mu}}^{\dagger}$ and $\hat{a}_{q_{\nu}}$. Taking expectation values of each term with the reference state $\Phi$, and using the fact that the expectation value of a normal ordered product vanishes gives
\begin{equation}
\begin{split}
\gamma^{p_{\mu}}_{q_{\nu}}	&= \braket{ \Phi | \hat{a}^{\dagger}_{p_{\mu}} \hat{a}_{q_{\nu}}	| 	\Phi } \\
                            &=		\braket{ \Phi |  \{ \hat{a}^{\dagger}_{p_{\mu}} \hat{a}_{q_{\nu}}	 \} | 	\Phi }+ \braket{ \Phi |  \wick{\c {\hat{a}_{p_{\mu}}^{\dagger} } \c { \hat{a}_{q_{\nu}} } } | 	\Phi }  \\
                            &= \braket{ \Phi |  \wick{\c {\hat{a}_{p_{\mu}}^{\dagger} } \c { \hat{a}_{q_{\nu}} } } | 	\Phi }  \\
\end{split}
\label{eqn:HoleContraction}
\end{equation}
where the definition of a $k$-particle reduced density matrix ($k$-RDM) is
\begin{equation}
    \gamma^{p_{\mu} q_{\nu} \cdots}_{r_{\sigma} s_{\tau} \cdots} = \bra{\Phi} \hat{a}_{p_{\mu}}^{\dagger} \hat{a}_{q_{\nu}}^{\dagger} \cdots \hat{a}_{s_{\tau}} \hat{a}_{r_{\sigma}} \ket{\Phi}
\end{equation}
Similarly, the $k$-hole reduced density matrix $\eta^{p_{\mu}}_{q_{\nu}}$ can be expressed as
\begin{equation}
\begin{split}
\eta^{p_{\mu}}_{q_{\nu}}    &= \braket{ \Phi |  \hat{a}_{q_{\nu}} \hat{a}^{\dagger}_{p_{\mu}}	| 	\Phi } \\
                            &= \braket{ \Phi |  \{ \hat{a}_{q_{\nu}} \hat{a}^{\dagger}_{p_{\mu}}		 \} | 	\Phi } + \braket{ \Phi |  \wick{\c {\hat{a}_{q_{\nu}} } \c { \hat{a}^{\dagger}_{p_{\mu}}  } } | 	\Phi } \\
                            &= \braket{ \Phi |  \wick{\c {\hat{a}_{q_{\nu}} } \c { \hat{a}^{\dagger}_{p_{\mu}}  } } | 	\Phi }
\end{split}
\end{equation}
$\eta^{p_{\mu}}_{q_{\nu}}$ is related to $\gamma^{p_{\mu}}_{q_{\nu}}$ through
\begin{equation}
\begin{split}
\eta^{p_{\mu}}_{q_{\nu}}    &= \braket{ \Phi |  \hat{a}_{q_{\nu}} \hat{a}^{\dagger}_{p_{\mu}}	| 	\Phi } \\
                            &= \braket{ \Phi |  \delta^{p_{\mu}}_{q_{\nu}} - \hat{a}^{\dagger}_{p_{\mu}} \hat{a}_{q_{\nu}}	| 	\Phi } \\
                            &= \delta^{p_{\mu}}_{q_{\nu}} - \gamma^{p_{\mu}}_{q_{\nu}} 
\end{split}
\end{equation}
One-body contractions give rise to one-particle ($\gamma^{p_{\mu}}_{q_{\nu}}$) or one-hole ($\eta^{p_{\mu}}_{q_{\nu}}$) RDMs. 
Equation \ref{eqn:HoleContraction} defines the conversion between an excitation operator normal ordered against the genuine vacuum and against the reference state.
\begin{equation}
\{ \hat{a}^{p_{\mu}}_{q_{\nu}}	 \} = \hat{a}^{p_{\mu}}_{q_{\nu}}	- \gamma^{p_{\mu}}_{q_{\nu}}
\label{eqn:SpinGNO1e}
\end{equation}
A similar approach can be used to bring two-electron operators into GNO.
\begin{equation}
\begin{split}
\{ \hat{a}^{p_{\mu} q_{\nu}}_{r_{\sigma} s_{\tau}}	 \} &=	\hat{a}^{p_{\mu} q_{\nu}}_{r_{\sigma} s_{\tau}}  - \gamma^{p_{\mu}}_{r_{\sigma}} \{ \hat{a}^{q_{\nu}}_{s_{\tau}} \} - \gamma^{q_{\nu}}_{s_{\tau}} \{ \hat{a}^{p_{\mu}}_{r_{\sigma}}\} + \gamma^{q_{\nu}}_{r_{\sigma}} \{ \hat{a}^{p_{\mu}}_{s_{\tau}}\} \\
&\quad + \gamma^{p_{\mu}}_{s_{\tau}} \{ \hat{a}^{q_{\nu}}_{r_{\sigma}} \} 
+ \gamma^{p_{\mu} }_{r_{\sigma} } \gamma^{q_{\nu}}_{s_{\tau}}  - \gamma^{p_{\mu} }_{s_{\tau} } \gamma^{q_{\nu}}_{r_{\sigma}}  + \lambda^{p_{\mu} q_{\nu}}_{r_{\sigma} s_{\tau}}
\end{split}
\label{eqn:SpinGNO2e}
\end{equation}
Here we have introduced the 2-cumulant $\lambda^{p_{\mu} q_{\nu}}_{r_{\sigma} s_{\tau}}$
\begin{equation}
  \lambda^{p_{\mu} q_{\nu}}_{r_{\sigma} s_{\tau}} = \gamma^{p_{\mu} q_{\nu}}_{r_{\sigma} s_{\tau}} - \gamma^{p_{\mu}}_{r_{\sigma}} \gamma^{q_{\nu}}_{s_{\tau}} + \gamma^{p_{\mu}}_{s_{\tau}} \gamma^{q_{\nu}}_{r_{\sigma}} = \wick{ \c1 {\hat{a}_{p_{\mu}}^{\dagger}}  \c2 {\hat{a}_{q_{\nu}}^{\dagger}} \c1{\hat{a}_{s_{\tau}}} \c2{\hat{a}_{r_{\sigma}}} }
\end{equation}
In general, the $k$-cumulant is written as\cite{kutzelniggCumulantExpansionReduced1999}
\begin{equation}
    \lambda^{p_{\mu} q_{\nu} \cdots}_{r_{\sigma} s_{\tau} \cdots} = \wick{ \c1 {\hat{a}_{p_{\mu}}^{\dagger}}  \c2 {\hat{a}_{q_{\nu}}^{\dagger}} \c3 {\cdots} \c3 {\cdots} \c1{\hat{a}_{s_{\tau}}} \c2{\hat{a}_{r_{\sigma}}} } 
\end{equation}
The $k$-cumulant can be understood as the fully connected component of the $k$-RDM.\cite{kongNovelInterpretationReduced2011,hanauerMeaningMagnitudeReduced2012} \\
In spin free formulations, one often works with unitary group generators of the form
\begin{equation}
\hat{E}^{pq ... }_{rs ...} = \sum_{\mu, \nu,... \in \{ \alpha, \beta\} }\hat{a}^{\dagger}_{p_{\mu}} \hat{a}^{\dagger}_{q_{\nu}} ... \hat{a}_{s_{\nu}} \hat{a}_{r_{\mu}}
\end{equation}
We can analogously define spin free operators in GNO by summing equations \ref{eqn:SpinGNO1e} and \ref{eqn:SpinGNO2e} over relevant spin indices.
\begin{equation}
\{ \hat{E}^{p}_{q}	 \} = \hat{E}^{p}_{q}	- \Gamma^{p}_{q}
\label{eqn:SFGNO1e}
\end{equation}
$\Gamma^{pq ... }_{rs ...}$ is the spin-free $k$-RDM given by
\begin{equation}
\begin{split}
    \Gamma^{pq ... }_{rs ...}   &= \braket{\Phi | \hat{E}^{pq ... }_{rs...} | \Phi} \\
                                &= \sum_{\mu, \nu,... \in \{ \alpha, \beta\} } \braket{\Phi | \hat{a}^{\dagger}_{p_{\mu}} \hat{a}^{\dagger}_{q_{\nu}} ... \hat{a}_{s_{\nu}} \hat{a}_{r_{\mu}} | \Phi} \\
                                &= \sum_{\mu, \nu,... \in \{ \alpha, \beta\} } \gamma^{p_{\mu} q_{\nu} \cdots}_{r_{\mu} s_{\nu} \cdots}
\end{split}
\end{equation}
A similar approach can be used to bring two-electron operators into GNO.
\begin{equation}
\begin{split}
\{ \hat{E}^{pq}_{rs}	 \} &=	\hat{E}^{pq}_{rs}  - \frac{1}{2}\Gamma^{p}_{r} \{ \hat{E}^{q}_{s} \} - \frac{1}{2}\Gamma^{q}_{s}  \{ \hat{E}^{p}_{r}\} + \frac{1}{2}\Gamma^{q}_{r} \{ \hat{E}^{p}_{s}\} \\
&\quad + \frac{1}{2}\Gamma^{p}_{s} \{ \hat{E}^{q}_{r} \} + \Gamma^{p}_{r}\Gamma^{q}_{s} -  \frac{1}{2}\Gamma^{p}_{s}\Gamma^{q}_{r} + \Lambda^{p q}_{r s}
\end{split}
\label{eqn:SFGNO2e}
\end{equation}
Bringing spin-free operators into GNO with respect to a given reference state can be interpreted as bringing the corresponding spin-orbital operators into GNO with respect to a $M_S$-averaged spin ensemble.\cite{kutzelniggNormalOrderExtended1997,kutzelniggIrreducibleBrillouinConditions2002,kutzelniggGeneralizedNormalOrdering2007,shamasundarCumulantDecompositionReduced2009,kutzelniggSpinfreeFormulationReduced2010} We provide a proof of this in Appendix \ref{Appendix:SpinEnsemble}.

\subsection{Spin-Ensemble Approach}
The normal ordered molecular Hamiltonian can be expressed in a spin-free way as
\begin{equation}
\hat{H}_{N} = E_0 + \sum_{pq} f^{q}_{p} \{ E^{p}_{q} \} + \frac{1}{2} \sum_{pqrs} g^{rs}_{pq} \{ E^{pq}_{rs} \}
\end{equation}
where we have defined the following:
\begin{align}
E_0 &= \sum_{pq} h^{q}_{p}\Gamma^{p}_{q} + \frac{1}{2} \sum_{pqrs} g^{qs}_{pr} \Gamma^{pr}_{qs}\\
f^{q}_{p} &= h^{q}_{p} + \sum_{rs} (g^{qs}_{pr} - \frac{1}{2} g^{qs}_{rp}) \Gamma^{r}_{s} \\
h^{q}_{p}   &=  \braket{q | \hat{h} | p }\\
g^{qs}_{pr} &= \braket{q s | p r} \equiv    \braket{q s | r_{12}^{-1} | p r}
\end{align}
$\hat{h}$ is the one-electron operator in the molecular Hamiltonian. While the equations are spin-free, many automated code generators work in a spin-orbital basis. Therefore, there is a need to convert between spin-orbital and spin-free quantities. This problem has been previously tackled, leading to a set of \emph{spin-replacement rules}\cite{shamasundarCumulantDecompositionReduced2009,kutzelniggSpinfreeFormulationReduced2010}. These rules relate the matrix elements in spin-orbital basis to the matrix elements in a spin-free basis. The rules are as follows: \\
1. For every Hamiltonian matrix element ($f^{q}_{p}$, $g^{qs}_{pr}$) or cluster amplitude $t^{rs \cdots}_{pq \cdots}$,
\begin{alignat}{2}
\Omega^{q_\alpha}_{p_\alpha} &= \Omega^{q_\beta}_{p_\beta} &&=  \Omega^{q}_{p}\\
\Omega^{q_\alpha s_\alpha}_{p_\alpha r_\alpha} &= \Omega^{q_\beta s_\beta}_{p_\beta r_\beta} &&= \Omega^{qs}_{pr} - \Omega^{qs}_{rp}\\
\Omega^{q_\alpha s_\beta}_{p_\alpha r_\beta} &= \Omega^{q_\beta s_\alpha}_{p_\beta r_\alpha} &&= \Omega^{qs}_{pr}\\
\Omega^{q_\alpha s_\beta}_{p_\beta r_\alpha} &= \Omega^{q_\beta s_\alpha}_{p_\alpha r_\beta} &&= - \Omega^{qs}_{rp}
\end{alignat}
2. For every 1-RDM (hole or particle) or $k$-cumulant,
\begin{alignat}{2}
\lambda^{p_{\alpha}}_{q_{\alpha}} &= \lambda^{P_{\beta}}_{Q_{\beta}} &&= \frac{1}{2} \lambda^{p}_{q}\\
\lambda^{p_{\alpha} q_{\alpha}}_{r_{\alpha} s_{\alpha}} &= \lambda^{p_{\beta} q_{\beta}}_{r_{\beta} s_{\beta}}  &&= \frac{1}{6} (\Lambda^{pq}_{rs} - \Lambda^{pq}_{sr})\\
\lambda^{p_{\alpha} q_{\beta}}_{r_{\alpha} s_{\beta}} &= \lambda^{p_{\alpha} q_{\beta}}_{r_{\alpha} s_{\beta}}  &&= \frac{1}{6} (2\Lambda^{pq}_{rs} + \Lambda^{pq}_{sr})\\
\lambda^{p_{\alpha} q_{\beta}}_{r_{\beta} s_{\alpha}} &= \lambda^{p_{\alpha} q_{\beta}}_{r_{\beta} s_{\alpha}}  &&= - \frac{1}{6} (\Lambda^{pq}_{rs} + 2\Lambda^{pq}_{sr})
\end{alignat}
These rules have previously been successfully employed in spin-free implementations of several multireference methods\cite{dattaStatespecificPartiallyInternally2011,liSpinfreeFormulationMultireference2021}. These spin-replacement rules have previously been used to spin-adapt many-body residuals.\cite{dattaStatespecificPartiallyInternally2011,liSpinfreeFormulationMultireference2021} In Appendix \ref{Appendix:SpinEnsemble}, we show that they can also be directly applied to coupled cluster equations.

\subsection{Wavefunction Ansatz}
In this work, we employ the GNO exponential wavefunction ansatz:
\begin{equation}
\ket{\Psi} = \{ e^{\hat{T}} \} \ket{\Phi}
\end{equation}
where the cluster operator $\hat{T}$ is of the form
\begin{equation}
\hat{T} = \sum_{pq} t^{q}_{p} \hat{E}^{p}_{q} + \frac{1}{2}\sum_{pqrs} t^{rs}_{pq}\hat{E}^{pq}_{rs} + \cdots
\end{equation}
The coupled cluster wavefunction $\ket{\Psi}$ is found by applying the generalised normal ordered exponential operator $\{ e^{\hat{T}} \}$ onto the multi-determinant wavefunction $\ket{\Phi}$ as first proposed by Lindgren.\cite{lindgrenCoupledclusterApproachManybody1978}  
In the limit where there are only core and virtual orbitals, the generalised normal ordered exponential operator reduces to the standard exponential operator used in single-reference coupled cluster.
A key advantage in using a normal ordered exponential is the simplification of the resulting working equations. As a corollary of Wick's theorem, operators within the same normal order do not contract with each other. Since all $\hat{T}$ in the exponential are within the same normal order, there will be no equations that involve pure $\wick{\c {\hat{T} } \c { \hat{T} } } $ contractions. The lack of contractions between cluster operators $\hat{T}$ also implies that the working equations have a finite Taylor expansion in order of $\hat{T}$. Beyond simplifying the resultant equations, the GNO formalism is also physically motivated as excitations parameterise independent correlation processes. \\
In this work, we will only use singles and doubles excitation operators such that
\begin{equation}
\hat{T} = \sum_{\substack{p \in \mathbb{A} \cup \mathbb{V}\\
                  q \in \mathbb{C} \cup \mathbb{A}\\
                  }} t^{q}_{p} \hat{E}^{p}_{q} + \frac{1}{2}\sum_{\substack{p,q \in \mathbb{A} \cup \mathbb{V}\\
                  r,s \in \mathbb{C} \cup \mathbb{A}\\
                  }} t^{rs}_{pq}\hat{E}^{pq}_{rs}
\end{equation}
We can group the various possible excitations into twelve \emph{excitation classes} (Table \ref{tab:ExcitationClasses}). For example, $\mathbb{CA} \rightarrow \mathbb{AV}$ represents a double excitation from a core and an active orbital into an active and a virtual orbital. As noted by Janssen and Schaefer\cite{janssenAutomatedSolutionSecond1991}, \emph{spectator} excitations are required in spin-free approaches to span the whole excitation space. We therefore allow for all spectator excitations.

\begin{table}[h!]
\renewcommand{\arraystretch}{1.5}
\setlength{\tabcolsep}{12pt}
\begin{tabular}{ c c c }
\hline
\hline
$\mathbb{C} \rightarrow \mathbb{A}$ & $\mathbb{CC} \rightarrow \mathbb{AA}$ &   $\mathbb{CA} \rightarrow \mathbb{VV}$\\
$\mathbb{C} \rightarrow \mathbb{V}$ & $\mathbb{CC} \rightarrow \mathbb{AV}$ &   $\mathbb{AA} \rightarrow \mathbb{AA}$\\
$\mathbb{A} \rightarrow \mathbb{A}$ & $\mathbb{CA} \rightarrow \mathbb{AA}$ &   $\mathbb{AA} \rightarrow \mathbb{AV}$ \\
$\mathbb{A} \rightarrow \mathbb{V}$ & $\mathbb{CA} \rightarrow \mathbb{AV}$ &   $\mathbb{AA} \rightarrow \mathbb{VV}$ \\
\hline
\hline
\end{tabular}
\caption{Excitation classes used in this work.}
\label{tab:ExcitationClasses}
\end{table}

\subsection{Working Equations}
\noindent
We begin from the Schrödinger equation
\begin{equation}
    \hat{H} \{ e^{\hat{T}} \} \ket{\Phi}    =      E \{ e^{\hat{T}} \} \ket{\Phi}
\label{eqn:Schrödinger}
\end{equation}
To arrive at working equations, one left-projects equation \ref{eqn:Schrödinger} with $\bra{\Phi}$ and $\bra{\Phi} \hat{\tau}_{\mu}^{\dagger}$ for the energy and residual equations, respectively. $\hat{\tau}$ denotes an arbitrary excitation operator.
The energy equation is given by
\begin{equation}
E = \bra{\Phi} \hat{H} \{ e^{\hat{T}} \} \ket{\Phi}
\label{eqn:Energy}
\end{equation}
and the residual equation is
\begin{equation}
\begin{split}
R_{\mu} &= \bra{\Phi} \hat{\tau}_{\mu}^{\dagger} \hat{H} \{ e^{\hat{T}} \} \ket{\Phi} - \bra{\Phi} \hat{\tau}_{\mu}^{\dagger} \{ e^{\hat{T}} \} \ket{\Phi}  \bra{\Phi} \hat{H} \{ e^{\hat{T}} \} \ket{\Phi} \\
&= \bra{\Phi}  \wick{\c1 {\hat{\tau}_{\mu}^{\dagger} } \c1 {\hat{H} }  \hspace{-4.8pt}\c2{\vphantom{{\hat{H} }}} \hspace{3.3pt} \c2 {\{ e^{\hat{T}} \}} } \ket{\Phi} \\ 
&= \bra{\Phi} \hat{\tau}_{\mu}^{\dagger} \hat{H} \{ e^{\hat{T}} \} \ket{\Phi}_{c} = 0
\end{split}
\label{eqn:Residual}
\end{equation}
In the last line, we introduced the subscript `c' to indicate that we are only taking the fully connected terms in the bra-ket $\bra{\Phi} \hat{\tau}_{\mu}^{\dagger} \hat{H} \{ e^{\hat{T}} \} \ket{\Phi}$. The GNO formalism ensures that all the working equations are fully connected. Evaluation of the energies and residuals involves only additive-separable quantities\cite{kutzelniggTheoryElectronCorrelation2003,misiewiczReducedDensityMatrix2020}, therefore ensuring size-extensivity. Detailed derivations of these equations are provided in Appendix \ref{Appendix:EquationsDerivation}.
These equations were initially proposed by Mukherjee and co-workers\cite{mahapatraStateSpecificMultiReferenceCoupled1998,sinhaGeneralizedAntisymmetricOrdered2013}, albeit without computer implementation to the best of the authors' knowledge. More recently, however, Mukherjee and co-workers implemented a version of these equations to correlate a single CSF (UGA-OSCC).\cite{senInclusionOrbitalRelaxation2018,chakravartiReappraisalNormalOrdered2021} \\

\subsubsection{Truncation of the exponential operator}
For our equations to be computationally tractable, we truncate the exponential operator at second order in its Taylor expansion. The energy and residual equations, therefore, read
\begin{equation}
E = \bra{\Phi} \hat{H} \{ 1 + \hat{T} + \frac{1}{2} \hat{T}^{2} \} \ket{\Phi}
\label{eqn:TruncatedEnergy}
\end{equation}
\begin{equation}
R_{\mu} = \bra{\Phi} \hat{\tau}_{\mu}^{\dagger} \hat{H} \{ 1 + \hat{T} + \frac{1}{2} \hat{T}^{2} \} \ket{\Phi}_{c} = 0
\label{eqn:TruncatedResidual}
\end{equation}
Higher order terms in $\hat{T}$ are expected to decrease in significance when a good reference is chosen and the cluster amplitudes are therefore small. This has been demonstrated numerically in the context of ic-MRCC\cite{evangelistaOrbitalinvariantInternallyContracted2011} where it was found that third and higher order terms could be neglected without significant loss of accuracy.

\subsubsection{Truncation of cumulant rank}
For a large active space, the energy and residual equations can contain high order cumulants. A $k$-cumulant is represented by a $k$-dimensional tensor, with $N^{k}$ elements, $N$ being the number of active orbitals. In principle, given an active space spanning $N$ active orbitals, cumulants of order up to $2N$ can be non-zero. This unfavourable scaling with active space size makes evaluating these equations very expensive if all possible cumulants are used. In our work, we retain all contributions up to 4-cumulants and discard higher-body terms. \\

Due to the connectedness of the GNOCC equations, size-extensivity is retained at every truncation level, in both the cluster amplitudes and cumulant order.

\subsection{Redundancy}

The amplitude equations are found by projecting the Schrödinger equation onto the excitation manifold. However, the residual equations $R_{\mu} = \braket{\Phi | \hat{\tau}_{\mu}^{\dagger} \hat{H} \{ e^{\hat{T}} \} | \Phi}_{c}$ and $R_{\nu} = \braket{\Phi | \hat{\tau}_{\nu}^{\dagger} \hat{H} \{ e^{\hat{T}} \} | \Phi}_{c}$ can be linearly dependent. This occurs because there can be multiple excitations leading to the same excited state, which leads to a singular overlap matrix 
\begin{equation}
    S_{\mu \nu} = \braket{ \Phi | \hat{\tau}_{\mu}^{\dagger} \hat{\tau}_{\nu} | \Phi }
\end{equation}
where $\hat{\tau}_{\mu}$ and $\hat{\tau}_{\nu}$ are cluster excitation operators.
To resolve such linear dependencies, it is typical to canonically orthogonalise the residual equations. This scheme has previously been employed in ic-MRCC implementations. We will briefly review the method and discuss the possibility of size-inconsistency using this approach.

\subsubsection{Canonical orthogonalisation of excitations}
We review the use of canonical orthogonalisation to remove redundant excitations. We begin with the $n \times n$ overlap matrix $S_{\mu \nu} = \braket{ \Phi | \hat{\tau}_{\mu}^{\dagger} \hat{\tau}_{\nu} | \Phi } $ of rank $m \leq n$, where $n$ is the number of cluster operators.
In a canonical orthogonalisation approach, one seeks a $n \times m$ transformation matrix $\boldsymbol{X}$ with matrix elements $X_{iI}$ such that 
\begin{equation}
    \boldsymbol{X}^{\dagger} \boldsymbol{S} \boldsymbol{X} = \sum_{\mu \nu}^{n}  X^{\dagger}_{i\mu} S_{\mu \nu} X_{\nu j} = \delta_{ij} = \boldsymbol{I}_{m \times m}
\end{equation}
Since the overlap matrix $\boldsymbol{S}$ is positive semidefinite, it can be diagonalised by some unitary matrix $\boldsymbol{U}$:
\begin{equation}
    \boldsymbol{U}^{\dagger} \boldsymbol{S} \boldsymbol{U} = \boldsymbol{\Sigma}
\end{equation}
$\boldsymbol{\Sigma}$ is a diagonal matrix with diagonal elements $\sigma_{n}$. 
The $n-m$ eigenvalues $\sigma_{n}$ smaller than a given numerical threshold $\epsilon$ are discarded, and the resulting $m \times m$ diagonal matrix is denoted by $\boldsymbol{\tilde{\Sigma}}$.
Similarly, $\boldsymbol{\tilde{U}}$ is the $n \times m$ matrix with the corresponding retained eigenvectors. The relationship between $\boldsymbol{\tilde{U}}$, $\boldsymbol{S}$, and $\boldsymbol{\tilde{\Sigma}}$ is given by
\begin{equation}
    \boldsymbol{\tilde{U}}^{\dagger} \boldsymbol{S} \boldsymbol{\tilde{U}} = \boldsymbol{\tilde{\Sigma}}
\end{equation}
Multiplying on the left and right on both sides of the equation by $\boldsymbol{\tilde{\Sigma}}^{-\frac{1}{2}}$, gives
\begin{equation}
    \boldsymbol{\tilde{\Sigma}}^{-\frac{1}{2}}\boldsymbol{\tilde{U}}^{\dagger} \boldsymbol{S} \boldsymbol{\tilde{U}} \boldsymbol{\tilde{\Sigma}}^{-\frac{1}{2}} = \boldsymbol{I}
\end{equation}
The transformation matrix $\boldsymbol{X}$ is therefore given by
\begin{equation}
    \boldsymbol{X} = \boldsymbol{\tilde{U}} \boldsymbol{\tilde{\Sigma}}^{-\frac{1}{2}}
\end{equation}
The use of canonical orthogonalisation has the benefit that it is orbital-invariant. However, we shall demonstrate that this approach can lead to size-inconsistency when the excitations contain spectators.

\subsubsection{Spectators and size-inconsistency}

The use of a non-redundant set of excitations given by the canonical orthogonalisation approach can also result in the inclusion of unwanted excitation terms. Consider the two-electron excitation operator $\hat{E}_{i t}^{t a}$ acting on a system consisting of subsystems $A$ and $B$ that are infinitely far apart. We assume that the orbitals corresponding to the labels $i,t,a$ are localised on either $A$ or $B$. The excitation with a core to virtual excitation localised on $A$, $\hat{E}_{i_A t}^{t a_A}$ will contain two components, $\hat{E}_{i_A t_A}^{t_A a_B}$ and $\hat{E}_{i_A t_B}^{t_B a_A}$ (equation \ref{eqn:ExcitationDecomposition}).
\begin{equation}
\hat{E}_{i_A t}^{t a_A} \rightarrow \hat{E}_{i_A t_A}^{t_A a_A}, \hat{E}_{i_A t_B}^{t_B a_A}
\label{eqn:ExcitationDecomposition}
\end{equation}
We can construct the overlap matrix in the basis of $\hat{E}_{i_A t_A}^{t_A a_A}$ and $\hat{E}_{i_A t_B}^{t_B a_A}$. Since the only orbitals that are changed are spectators (which do not change the reference state), both excitations are equivalent. The overlap matrix is therefore given by

\begin{equation}
\begin{pmatrix}
\braket{\Phi | \hat{E}_{i_A t_A}^{t_A a_A \dagger} \hat{E}_{i_A t_A}^{t_A a_A} | \Phi} & \braket{\Phi | \hat{E}_{i_A t_A}^{t_A a_A \dagger} \hat{E}_{i_A t_B}^{t_B a_A} | \Phi}\\
\braket{\Phi | \hat{E}_{i_A t_B}^{t_B a_A \dagger} \hat{E}_{i_A t_A}^{t_A a_A} | \Phi} & \braket{\Phi | \hat{E}_{i_A t_B}^{t_B a_A \dagger} \hat{E}_{i_A t_B}^{t_B a_A} | \Phi}\\
\end{pmatrix}
= 
\begin{pmatrix}
1 & 1\\
1 & 1
\end{pmatrix}
\end{equation}
The non-null eigenvector is therefore given by
\begin{equation}
\hat{E} = \frac{1}{\sqrt{2}} \Big( \hat{E}_{i_A t_A}^{t_A a_A} + \hat{E}_{i_A t_B}^{t_B a_A} \Big)
\end{equation}
Due to the equivalence of their action on the reference, both excitations will always be coupled in the non-null eigenvector. As argued in a previous work,\cite{gunasekeraMultireferenceCoupledCluster2024} the spurious term $\hat{E}_{i_A t_B}^{t_B a_A}$ leads to size-inconsistent energies as the excitation basis fails to be additively separable. Therefore, a procedure to remove the unwanted $\hat{E}_{i_A t_B}^{t_B a_A}$ term is required.\\
A possible solution presents itself from the observation that in the Hamiltonian, the operators $\hat{E}_{i_A t_A}^{t_A a_A}$ and $\hat{E}_{i_A t_B}^{t_B a_A}$ come with coefficients $g^{i_A t_A}_{t_A a_A}$ and $g^{i_A t_B}_{t_B a_A}$, respectively. In the case where $A$ and $B$ are infinitely far apart, only $g^{i_A t_B}_{t_B a_A} \rightarrow 0$. This suggests that the selection of operators should be determined by their coefficients in the Hamiltonian expression. In other words, we want to pick excitations which will bring the reference into the \emph{first order interacting space} (FOIS). We will describe a procedure for doing so in the following section.

\subsubsection{Expressing the FOIS}
We shall now derive a transformation matrix that converts a non-orthogonal and potentially linearly dependent set of excitations into a set of orthogonalised, linearly independent excitations that excite the reference state into the FOIS.\\
Expressing the Hamiltonian operator in the general form 
\begin{equation}
    \hat{H} = \sum_{\mu} h_{\mu} \hat{\tau}_{\mu}
\end{equation}
we can construct a weighted overlap matrix 
\begin{equation}
    \tilde{S}_{\mu \nu} = h_{\mu} \braket{ \Phi | \hat{\tau}_{\mu}^{\dagger} \hat{\tau}_{\nu} | \Phi } h_{\nu} = h_{\mu} S_{\mu \nu} h_{\nu}
\label{eqn:WeightedOverlap}
\end{equation}
The weighted overlap matrix can be diagonalised through a canonical orthogonalisation approach such that
\begin{equation}
    \sum_{\mu \nu} \tilde{X}_{i \mu}^{\dagger} \tilde{S}_{\mu \nu} \tilde{X}_{\nu i} = \delta_{ij}
\end{equation}
Having found the canonical transformation matrix $\tilde{\bf{X}}$, the excitations leading to the FOIS, $\hat{\tau}^{\text{FOIS}}_{i}$ can be expressed as
\begin{equation}
    \hat{\tau}^{\text{FOIS}}_{i} = \sum_{\mu} \hat{\tau}_{\mu} h_{\mu} \tilde{X}_{\mu i} = \sum_{\mu} \hat{\tau}_{\mu} Y_{\mu i}
\end{equation}
where we have defined the transformation matrix $Y_{\mu i} = h_{\mu} \tilde{X}_{\mu i}$, which converts excitations into a set of orthogonalised excitations that generate the FOIS from the reference state. $Y_{\mu i}$ also satisfies
\begin{equation}
\begin{split}
\sum_{\mu \nu} Y_{i \mu}^{\dagger} \braket{\Phi | \hat{\tau}^{\dagger}_{\mu} \hat{\tau}_{\nu} | \Phi } Y_{\nu j} 
&= \sum_{\mu \nu} \tilde{X}_{i \mu}^{\dagger} h_{\mu} \braket{\Phi | \hat{\tau}^{\dagger}_{\mu} \hat{\tau}_{\nu} | \Phi } h_{\nu} \tilde{X}_{\nu j} \\
&= \sum_{\mu \nu} \tilde{X}_{i \mu}^{\dagger} \tilde{S}_{\mu \nu} \tilde{X}_{\nu j} \\
& = \delta_{ij}
\end{split}
\label{eqn:YOrthogonality}
\end{equation}
With this transformation scheme, spurious excitation terms at dissociation are excluded, ensuring size-consistency. However, this non-unitary transformation of the excitation basis implies the loss of orbital invariance. Since our proposed method is not orbital invariant, we require a method to specify an orbital basis for our reference functions. The selection of excitations leading to the FOIS requires orbitals that localise on either fragment in the dissociation limit. In this work, we elect to localise orbitals through the Pipek-Mezey scheme with Becke charges\cite{lehtolaPipekMezeyOrbital2014} to localise orbitals in the active space.\\
Once we have defined the transformation matrix $\bm{Y}$, we can determine the relationship between the cluster amplitudes in the linearly dependent and the linearly independent basis. Noting that the cluster operator, $\hat{T}$, can be expressed in the linearly dependent basis as
\begin{equation}
    \hat{T} = \sum_{\mu} t_{\mu} \hat{\tau}_{\mu}
\label{eqn:ClusterLinDep}
\end{equation}
or in the linearly independent basis as
\begin{equation}
    \hat{T} = \sum_{i \mu} \tilde{t}_{i} \hat{\tau}_{\mu} Y_{\mu i}
\label{eqn:ClusterLinIndep}
\end{equation}
where $t_{\mu}$ and $\tilde{t}_{i}$ are cluster amplitudes in the linearly dependent and linearly independent basis, respectively. Combining equations \ref{eqn:ClusterLinDep} and \ref{eqn:ClusterLinIndep} gives
\begin{equation}
    t_{\mu} = \sum_{i} Y_{\mu i} \tilde{t}_{i}
\label{eqn:ProjectorRequirement}
\end{equation}
This defines the relationship between cluster amplitudes in different bases.

\subsection{Amplitude update equations}
Here we detail the method by which we solve the residual equations. The residual equation in the linearly independent basis is given by
\begin{equation}
R_{i} = \sum_{\mu} Y_{i \mu}^{\dagger} \bra{\Phi} \hat{\tau}_{\mu}^{\dagger} \hat{H} \{ e^{\hat{T}} \} \ket{\Phi}_{c}
\end{equation}
This is found by projecting the Schrödinger equation onto the basis of linearly independent states spanning the FOIS. Left-multiplying both sides of the equation by $Y_{\sigma i}$ and summing over $i$ gives
\begin{equation}
\tilde{R}_{\sigma} \equiv \sum_{i} Y_{\sigma i} R_{i} = \sum_{i \mu} Y_{\sigma i} Y_{i \mu}^{\dagger} \bra{\Phi} \hat{\tau}_{\mu}^{\dagger} \hat{H} \{ e^{\hat{T}} \} \ket{\Phi}_{c}
\end{equation}
We seek a change in cluster amplitude $\delta \hat{T} = \sum_{\nu} \hat{\tau}_{\nu} \delta t_{\nu}$ such that the change in $\tilde{R}_{\sigma}$ with respect to $\delta \hat{T}$ is zero. That is,
\begin{equation}
\delta \tilde{R}_{\sigma} = \sum_{i \mu} Y_{\sigma i} Y_{i \mu}^{\dagger}  \bra{\Phi} \hat{\tau}_{\mu}^{\dagger} \hat{H} \{ e^{\hat{T} + \delta \hat{T}} \} \ket{\Phi}_{c} = 0
\label{eqn:AmplitudeChange}
\end{equation}
We now introduce the approximation  $\{ e^{\hat{T} + \delta \hat{T}} \} \approx \{ e^{\hat{T} } + \delta \hat{T} \}$ to arrive at
\begin{equation}
\begin{split}
&\sum_{i \mu} Y_{\sigma i} Y_{i \mu}^{\dagger} \bra{\Phi} \hat{\tau}_{\mu}^{\dagger} \hat{H} \{ e^{\hat{T}} \} \ket{\Phi}_{c} \\
&\hspace{2cm}+ \sum_{i \mu \nu} Y_{\sigma i} Y_{i \mu}^{\dagger}  \bra{\Phi} \hat{\tau}_{\mu}^{\dagger} \hat{H} \hat{\tau}_{\nu} \ket{\Phi}_{c} \delta t_{\nu} = 0
\end{split}
\end{equation}
The first term is simply $\tilde{R}_{\sigma}$. To simplify the evaluation of the second term, we partition the Hamiltonian into a zeroth-order and first-order term $ \hat{H} = \hat{H}_{0} + \hat{H}_{1}$
and retain in $\bra{\Phi} \hat{\tau}_{\mu}^{\dagger} \hat{H} \hat{\tau}_{\nu} \ket{\Phi}_{c}$ only the terms containing $\hat{H}_{0}$. This gives the approximate update equation for the cluster amplitudes $t_{\nu}$.
\begin{equation}
    \tilde{R}_{\sigma} + \sum_{i \mu \nu} Y_{\sigma i} Y_{i \mu}^{\dagger}  \bra{\Phi} \hat{\tau}_{\mu}^{\dagger} \hat{H}_{0} \hat{\tau}_{\nu} \ket{\Phi}_{c} \delta t_{\nu} = 0
\label{eqn:MacroUpdate}
\end{equation}
In this work, we use the Dyall Hamiltonian\cite{dyallChoiceZerothorderHamiltonian1995} as $\hat{H}_{0}$, that is, 
\begin{equation}
    \hat{H}_{0} = \sum_{ij} f^{j}_{i} \hat{E}_{j}^{i} + \sum_{ab} f^{b}_{a} \hat{E}_{b}^{a} + \sum_{tu} f^{u}_{t} \hat{E}_{u}^{t} + \frac{1}{2} \sum_{tuvw} g^{vw}_{tu} \hat{E}_{vw}^{tu}
\end{equation}
$\delta t_{\nu}$ in equation \ref{eqn:MacroUpdate} is solved for iteratively in a separate routine (micro-iteration). In this procedure, the change in $\delta t_{\nu}$ at each step, $\delta^{(2)} t_{\nu}$, is given by
\begin{equation}
    \delta^{(2)} t_{\nu} = \frac{\tilde{R}_{\nu} + \sum_{i \mu \lambda} Y_{\sigma i} Y_{i \mu}^{\dagger}  \bra{\Phi} \hat{\tau}_{\mu}^{\dagger} \hat{H}_{0} \hat{\tau}_{\lambda} \ket{\Phi}_{c} \delta t_{\lambda}}{\Delta_{\nu}}
\end{equation}
where we define
\begin{equation}
    \Delta_{\nu} \equiv \Delta_{rs \cdots}^{pq \cdots} = f_{p}^{p} + f_{q}^{q} + \cdots - f_{r}^{r} - f_{s}^{s} - \cdots
\end{equation}
At convergence, $R_{\mu} + \sum_{\nu} \bra{\Phi} \hat{\tau}_{\mu}^{\dagger} \hat{H}_{0} \hat{\tau}_{\nu} \ket{\Phi}_{c} \delta t_{\nu} = 0$ and hence $\delta^{(2)} t_{\nu} = 0$. Since it is possible that $\Delta_{\nu} = 0$, leaving the micro-iteration amplitude update undefined, we apply a level shift $\eta$ to the denominator in practical implementation, such that
\begin{equation}
    \delta^{(2)} t_{\nu} = \frac{\tilde{R}_{\nu} + \sum_{i \mu \lambda} Y_{\sigma i} Y_{i \mu}^{\dagger}  \bra{\Phi} \hat{\tau}_{\mu}^{\dagger} \hat{H}_{0} \hat{\tau}_{\lambda} \ket{\Phi}_{c} \delta t_{\lambda}}{\Delta_{\nu} + \eta}
\end{equation}

\subsection{Amplitude Projection}
The resultant cluster amplitudes have to be projected to remove redundant components from the updated amplitudes so that the relationship between cluster amplitudes in different bases (equation \ref{eqn:ProjectorRequirement}) is upheld. A suitable projector matrix $\boldsymbol{P}$ with matrix elements $P_{\mu \sigma}$ is given by
\begin{equation}
    P_{\mu \sigma} = \sum_{i \mu \nu} Y_{\mu i} Y^{\dagger}_{i \nu} S_{\nu \sigma}
\label{eqn:Projector}
\end{equation}
The idempotency of $\bm{P}$ can be easily shown using equation \ref{eqn:YOrthogonality}. We can further examine the result of applying the projector onto the cluster amplitude in the linearly dependent basis, $t_{\sigma}$. Using equations \ref{eqn:ProjectorRequirement}, \ref{eqn:Projector}, and \ref{eqn:YOrthogonality}, we can express the projected amplitude, $\sum_{\sigma} P_{\mu \sigma} t_{\sigma}$, as
\begin{equation}
    \begin{split}
        \sum_{\sigma} P_{\mu \sigma} t_{\sigma} &=   \sum_{j \sigma} P_{\mu \sigma} Y_{\sigma j} \tilde{t}_{j} \\
                    &= \sum_{ij \nu \sigma} Y_{\mu i} Y^{\dagger}_{i \nu} S_{\nu \sigma} Y_{\sigma j} \tilde{t}_{j} \\ 
                    &=  \sum_{ij} Y_{\mu i} \delta_{ij} \tilde{t}_{j} \\ 
                    &=  \sum_{j} Y_{\mu j}\tilde{t}_{j} \\ 
                    &=  t_{\mu}
    \end{split}
\end{equation}
This shows that when equation \ref{eqn:ProjectorRequirement} is fulfilled, the projector leaves the cluster amplitude unchanged. This property, along with its idempotency, qualifies its use as a projector. We shall therefore use $\boldsymbol{P}$ to project out any redundant components from the cluster amplitudes. 

\subsection{Iterative cycle}
The coupled cluster equations are non-linear and are thus solved iteratively.
To improve convergence, we adopt a macro-micro iteration scheme. In each macro-iteration cycle,
\begin{enumerate}
    \item The energy $E = \bra{\Phi} \hat{H} \{ e^{\hat{T}} \} \ket{\Phi}$ and the residual $\boldsymbol{R}$ with elements $R_{\mu} = \bra{\Phi} \hat{\tau}_{\mu}^{\dagger} \hat{H} \{ e^{\hat{T}} \} \ket{\Phi}_{c}$ are evaluated using the current amplitudes \textbf{t}. The residual matrix is computed in the linearly dependent basis.
    \item The change in cluster amplitudes, $\delta t_{\nu}$, is computed through the micro-iterative cycle where equation \ref{eqn:MacroUpdate} is solved iteratively.
    \item The projector $\bm{P} = \bm{YY^{\dagger}S}$ is applied onto $\delta t_{\nu}$ so that the change in cluster amplitudes satisfy equation \ref{eqn:ProjectorRequirement}.
    \item The cluster amplitudes are updated, and this macro-iterative cycle repeats until convergence of the residual norm. 
\end{enumerate}
In both the macro- and micro-iteration cycles, convergence was accelerated by applying the Direct Inversion of the Iterative Subspace (DIIS)\cite{pulayConvergenceAccelerationIterative1980,pulayImprovedSCFConvergence1982} method.

\subsection{Decoupling of the overlap matrix}
The treatment of redundancies requires the construction and diagonalisation of the weighted overlap matrix (equation \ref{eqn:WeightedOverlap}), where $\hat{\tau}_{\mu}$ and $\hat{\tau}_{\nu}$ can be excitations from different excitation classes. 
Fortuitously, excitations from different excitation classes are often linearly independent. This is because excitations involving different numbers of core and virtual indices are orthogonal. Therefore, most linear dependencies come from excitations within the same excitation class. This allows us to decouple the overlap matrix into smaller sub-blocks that can be diagonalised separately.\\
The only exceptions to this in our work comes from the linear dependence between $\mathbb{P} \rightarrow \mathbb{Q}$ and $\mathbb{PA} \rightarrow \mathbb{AQ}$, where $\mathbb{P} \in \{ \mathbb{C} \cup \mathbb{A} \}$ and $\mathbb{Q} \in \{ \mathbb{A} \cup \mathbb{V} \}$. All possible non-zero overlaps between excitation classes and the analytical expressions for their corresponding overlap matrices are detailed in Appendix \ref{Appendix:RedundancyOverlap}. We express these overlap matrices purely in terms of spin-free quantities so no further spin-adaptation is required.


\subsection{Scaling}
The working equations in this work were generated through the use of an automated equation generation package, \texttt{Wick\&D}\cite{evangelistaAutomaticDerivationManybody2022}. Each term in the coupled cluster equation is written as a tensor contraction using the \texttt{opt\_einsum} package\cite{smithOpt_einsumPythonPackage2018}. We can determine the scaling by analysing the tensor contraction path in each term and identifying the bottlenecks in both computation and memory use. Recently, Feldman and Reiher\cite{feldmannRenormalizedInternallyContracted2024} performed a similar analysis on their automatically generated renormalised ic-MRCC (ric-MRCC) equations.\\
We shall label the number of core orbitals as $n_{C}$, the number of active orbitals as $n_{A}$, and the number of virtual orbitals as $n_{V}$. If we assume that $n_{A} < n_{C} << n_{V}$, the most expensive step scales as $n_{C}^{2} n_{V}^{4}$. This is due to the residual equation for the  $\mathbb{CC} \rightarrow \mathbb{VV}$ amplitude and is therefore unsurprising that it shares the same scaling as single-reference CCSD. If $n_{C} < n_{A} << n_{V}$ instead, the scaling will be $n_{A}^{2} n_{V}^{4}$ due to the residual equation for the $\mathbb{AA} \rightarrow \mathbb{VV}$ amplitude. However, if $n_{C} \sim n_{V} << n_{A}$, the computational scaling becomes $n_{A}^{11}$ due to the presence of contraction terms between high-order cumulants. Therefore, it is imperative that the number of active orbitals should be small compared to the number of virtual orbitals for the method to have scaling comparable to that of CCSD. \\
The other bottleneck comes from memory use. The redundancy handling procedure requires the construction of an overlap matrix in the basis of excitations. In the case where $n_{A} < n_{C} << n_{V}$, the overlap matrix for the $\mathbb{AA} \rightarrow \mathbb{VV}$ excitation class will contain $n_{A}^{4} n_{V}^{4}$ elements. In the limit of $n_{C} \sim n_{V} << n_{A}$, the most memory-intensive step is the construction of the overlap matrix for the $\mathbb{AA} \rightarrow \mathbb{VV}$ excitation class, along with the evaluation of the 4-cumulant, both of which scales as $n_{A}^{8}$.
The unfavourable scaling with respect to the number of active orbitals suggests that we are currently limited to a relatively small size of active space. For example, a modest active space of 12 orbitals, or 24 spin-orbitals, results in a 4-cumulant that requires over 800 GB of memory.

\section{Comparison to related approaches}
The GNOCC formalism shall now be contrasted with several state-specific multi-determinantal coupled cluster theories explored in earlier literature. These theories bear similarities to ours either by way of choice of ansätze or working equations.

\subsection{UGA-OSCC and NOECC}
Mukherjee's UGA-OSCC\cite{senInclusionOrbitalRelaxation2018,chakravartiReappraisalNormalOrdered2021} method, the NOECC\cite{gunasekeraMultireferenceCoupledCluster2024} method, and GNOCC all employ a normal-ordered exponential ansatz. In GNOCC, normal ordering of the exponential operator is defined with respect to the given reference, whereas in UGA-OSCC and NOECC, it is defined with respect to a closed-shell vacuum where only the core orbitals are doubly occupied and with all other orbitals unoccupied. 
The UGA-OSCC and NOECC methods are very similar to each other, differing mainly in the approximations used to derive the working equations. 
In UGA-OSCC, the Schrödinger equation (equation \ref{eqn:Schrödinger}) is recast as the Bloch equation\cite{blochTheoriePerturbationsEtats1958}
\begin{equation}
    \hat{H} \{ e^{\hat{T}} \}_{\text{cs}} \ket{\Phi}    =      \{ e^{\hat{T}} \}_{\text{cs}} \hat{H}_{\text{eff}} \ket{\Phi}
\label{eqn:Bloch}
\end{equation}
where $\hat{H}_{\text{eff}}$ is an effective Hamiltonian such that $\hat{H}_{\text{eff}}  \ket{\Phi} = E \ket{\Phi}$. The use of the subscript ``cs'' indicates that the normal ordering is with respect to the closed-shell vacuum.  Successive application of Wick's theorem results in an infinite series, which is truncated to give the various forms of the UGA-OSCC method. \\
In NOECC, a more direct approach was taken to arrive at working equations. The energy equation results from left-projecting the Schrödinger equation with the reference state $\ket{\Phi}$, giving
\begin{equation}
    E = \bra{\Phi} \hat{H} \{ e^{\hat{T}} \}_{\text{cs}} \ket{\Phi}
\label{eqn:NOECCEnergy}
\end{equation}
This is similar to the energy equation used for GNOCC (equation \ref{eqn:Energy}), differing only in the definition of the normal ordering.
The residual equation is found by left-projecting onto excited states and is given by
\begin{equation}
R_{\mu} = \bra{\Phi} \hat{\tau}_{\mu}^{\dagger} \hat{H} \{ e^{\hat{T}} \}_{\text{cs}} \ket{\Phi} - \bra{\Phi} \hat{\tau}_{\mu}^{\dagger} \{ e^{\hat{T}} \}_{\text{cs}} \ket{\Phi}  \bra{\Phi} \hat{H} \{ e^{\hat{T}} \}_{\text{cs}} \ket{\Phi} \\
\label{eqn:NOECCResidual}
\end{equation}
This equation differs from the first line of equation \ref{eqn:Residual} only in the definition of the normal ordering. However, it is not straightforward to simplify this expression into the fully connected form used in this work because the normal ordering is defined only with respect to the closed-shell vacuum. In fact, truncating equation \ref{eqn:NOECCResidual} to a lower order in cluster amplitudes leads to the appearance of disconnected terms. The NOECC residual equation is only fully connected when all terms are included up to the order at which the residual equation terminates. This is contrasted with the GNOCC working equations which remains connected at every level of truncation of cluster amplitudes.\\
Apart from differences in the ansatz and working equations, GNOCC differentiates itself in its handling of redundancies in the excitation basis. In UGA-OSCC, a linearly independent set of excitation operators is chosen \emph{a priori} similar to that of UGA approaches\cite{liUnitarygroupbasedOpenshellCoupledcluster1998}. As such, no redundancies exist between these operators. Meanwhile, redundancies were not explicitly handled in NOECC as the residual equations were found to converge with appropriate numerical techniques.\\
For GNOCC, we allow for the use of any arbitrary spin eigenfunctions, which in turn makes the \emph{a priori} generation of a linearly independent set of excitation operators challenging. We therefore seek out non-redundant excitations through diagonalisation of a weighted overlap matrix in the spirit of internally-contracted methods.

\subsection{ic-MRCC}
It is instructive to compare our formalism to that of ic-MRCC. A key difference between GNOCC and ic-MRCC lies in the wave operator. In GNOCC, the wave operator is a normal-ordered exponential $\{ e^{\hat{T}} \}$, where $\hat{T}$ includes all possible excitations from the space of core and active orbitals to the space of active and virtual orbitals. In ic-MRCC, the wave operator is $e^{\hat{T}} $, and purely active-to-active excitations are excluded from $\hat{T}$. The differences in wave operator will lead to differing formulations of working equations.\\
In ic-MRCC approaches, connected equations are found by using the similarity transformed Hamiltonian
\begin{equation}
    \bar{H} = e^{-\hat{T}} \hat{H} e^{\hat{T}} = \hat{H} + [\hat{H}, \hat{T}] + \frac{1}{2} [[\hat{H}, \hat{T}], \hat{T}] + \cdots
\end{equation}
which can be expressed as a sum of nested commutators through the BCH expansion. This is in contrast to GNOCC where the similarity transformation is not performed because the inverse of $\{ e^{\hat{T}} \}$ is complicated. In both cases, the resulting equations are all connected and, therefore, size-extensive. To better understand the difference between the two sets of equations, we can compare their respective energy equations by order in $\hat{T}$. The ic-MRCC energy equation up to second order in $\hat{T}$ is given by
\begin{equation}
    E^{\text{ic-MRCC}} = \braket{\Phi | \bar{H} | \Phi } = \braket{\Phi |  \hat{H} + [\hat{H}, \hat{T}] + \frac{1}{2} [[\hat{H}, \hat{T}], \hat{T}] | \Phi }
\label{eqn:icMRCCEnergy}
\end{equation}
Recalling that in ic-MRCC, no purely active-to-active excitations are present in the excitation operators $\hat{T}$. $\hat{T}$ must therefore contain either creation operators with virtual orbital labels or annihilation operators with core orbital labels. As a result,
\begin{equation}
    \bra{\Phi} \hat{T} = (\hat{T}^{\dagger} \ket{\Phi})^{\dagger} = 0
\end{equation}
because $\hat{T}^{\dagger}$ must either annihilate from a virtual orbital or create into a core orbital, both of which give a vanishing result. With this result, we can evaluate the expectation values of the commutators $[\hat{H}, \hat{T}]$ and $[[\hat{H}, \hat{T}], \hat{T}]$ by
\begin{equation}
    \begin{split}
        \braket{\Phi |  [\hat{H}, \hat{T}]  | \Phi } 
        &= \braket{\Phi |  \hat{H}\hat{T} -  \hat{T}\hat{H}  | \Phi } \\
        &= \braket{\Phi |  \hat{H}\hat{T}  | \Phi } \\
        &= \braket{\Phi |  \hat{H}\hat{T}  | \Phi }_{c} \\
    \end{split}
\end{equation}
and 
\begin{equation}
    \begin{split}
        \braket{\Phi |  [[\hat{H}, \hat{T}], \hat{T}]  | \Phi } 
        &= \braket{\Phi |  \hat{H}\hat{T}\hat{T}  | \Phi } \\
        &= \braket{\Phi |  \hat{H}\hat{T}^{2}  | \Phi }_{c} \\
    \end{split}
\end{equation}
By substituting these results into equation \ref{eqn:icMRCCEnergy}, we can express the ic-MRCC energy to second order in $\hat{T}$ as
\begin{equation}
    E^{\text{ic-MRCC}} = \braket{\Phi |  \hat{H} + \hat{H}\hat{T} + \frac{1}{2} \hat{H}\hat{T}^{2} | \Phi }_{c}
\end{equation}
which has the same form as the GNOCC energy equation, equation \ref{eqn:Energy}. However, even in the limit that purely active-to-active excitations are omitted from GNOCC, we note that the energy equations are still different because $\hat{T}^{2}$ is not in normal order. As such, terms involving $\wick{\c {\hat{T} } \c { \hat{T} } } $ contractions can occur in ic-MRCC. \\
This analysis motivates our truncation of the normal-ordered exponential at second order in $\hat{T}$ as it is analogous to the typical choice of truncating at the two-fold commutator in ic-MRCC methods.



\section{Results and Discussion}
All GNOCC calculations were performed using an in-house \texttt{PYTHON} implementation. The working equations were generated using the \texttt{Wick\&D} package\cite{evangelistaAutomaticDerivationManybody2022}. CAS and FCI calculations were performed using \texttt{PySCF}.\cite{sunPySCFPythonbasedSimulations2018,sunRecentDevelopmentsPySCF2020} In all calculations using a CASSCF reference, optimised orbitals from a CASSCF calculation are localised within the core, active, and virtual spaces, respectively. The localisation procedure used is the Pipek-Mezey localisation with Becke charges\cite{lehtolaPipekMezeyOrbital2014} and was performed with \texttt{PySCF}. The CASSCF wavefunction with localised orbitals is subsequently used as the reference for GNOCC calculations.

\subsection{He atom}
We shall use the helium atom as a simple example of a two-electron system. For a two-electron system, we expect GNOCC with singles and doubles excitations (GNOCCSD) to be exact because all possible determinants can be accessed through single and double excitations from the reference state. Therefore, comparison of GNOCCSD to full configuration interaction (FCI) for the helium atom provides an important test for the correctness of our implementation. 
The calculations were performed with a cc-pVTZ basis, and a CASSCF(2,2) calculation with the desired orbital/spin configuration was performed in all cases to provide the starting reference. 

\begin{table}[h!]
\renewcommand{\arraystretch}{1.5}
\resizebox{\columnwidth}{!}{\begin{tabular}{ c c c c c }
\hline
\hline
State                   &		GNOCCSD(2)		&		GNOCCSD(3)		&		GNOCCSD(4)		&		FCI	\\
\hline
$^{1}S$	($1s^{2}$)		&		-2.900510	    &		-2.900231       &		-2.900232       &     -2.900232                        \\
$^{3}S$	($1s2s$)		&	    -1.915080		&		-1.915080		&		-1.915086		&     -1.915086                          \\
$^{1}S$	($1s2s$)		&		-1.718102		&		-1.718752		&		-1.718293		&     -1.718293                          \\
$^{3}P$	($1s2p$)		&		-1.254174		&		-1.254174		&		-1.254206		&     -1.254206                         \\
$^{1}P$	($1s2p$)		&		-1.019298		&		-1.019298	    &		-1.019798		&     -1.019798                         \\
\hline
\hline
\end{tabular}}
\caption{Energies (in Hartrees) of the ground state and various excited states of Helium atom. GNOCCSD($k$) refers to a truncation at the $k$-cumulant. For example, we exclude all terms with $3$- and higher order cumulants in GNOCCSD($2$).}
\label{tab:HeliumEnergies}
\end{table}

We have tabulated (Table \ref{tab:HeliumEnergies}) the GNOCCSD energies for ground and low-lying excited states of the helium atom at varying levels of cumulant truncation. We shall denote GNOCCSD($k$) as GNOCCSD with up to, and including, the $k$-body cumulant. With all 5- and higher-body cumulants vanishing, GNOCCSD(4) should be exact, and this is demonstrated numerically by comparison with the FCI energies. Using different starting references, we were able to target various excited states for helium, including both the closed-shell and open-shell $^{1}S$ states, showing promise for state-specific applications. \\
By truncating GNOCCSD equations at different cumulant orders, we find that GNOCCSD(2) and GNOCCSD(3) generally give energies that are in good agreement with FCI. In fact, GNOCCSD(2) and GNOCCSD(3) give the same energies for all but the $^{3}S$ and $P$ states in Table \ref{tab:HeliumEnergies}. This is because in these cases, the 3-cumulant vanishes, and therefore GNOCCSD(2) and GNOCCSD(3) are equivalent. The reason for the 3-cumulant vanishing is because in these states, the active space is symmetric with respect to exchange between particles and holes\cite{hanauerMeaningMagnitudeReduced2012}. For the higher-lying excited states such as $^{1}S$	($1s2s$) and $^{1}P$ ($1s2p$), neglecting the 4-cumulant leads to energy errors of the order of $0.1$mE$_{\text{H}}$. This indicates that it will be prudent to include higher-order cumulants for the sake of numerical accuracy. In this work, we will use GNOCCSD(4) throughout.\\
It is also interesting to note that the corresponding triplet states, $^{3}S$	($1s2s$) and $^{3}P$ ($1s2p$), exhibit much smaller errors from cumulant truncation. We believe that this is related to the use of a $M_S$-averaged ensemble formalism where a triplet state is represented by an equally weighted linear combination of the three possible $M_S$ states. Since $M_S = \pm 1$ states can be represented by a single Slater determinant, they do not contribute to the cumulants. As such, the cumulant dependence of GNOCCSD for triplet calculations is smaller relative to that of the corresponding open-shell singlet, and this is reflected in the differences in error with cumulant truncation.


\subsection{Li$_2$}
The lithium dimer is a simple yet revealing test case for examining size consistency in open-shell coupled cluster approaches. We calculate the correlation energy of both the lowest-lying open-shell $^{3}\Sigma_{g}^{+}$ and $^{1}\Sigma_{g}^{+}$ states for a well-separated lithium dimer. For a size-consistent electronic structure method, the correlation energy of the dimer for either spin state should be exactly double that of the $^{2}S$ state of the lithium atom. Therefore, we define the size-consistency error, $\Delta E$ as
\begin{equation}
    \Delta E (\text{Li}_{2}) = 2E(^{2}\text{Li}) - E(\text{Li}_{2})
\end{equation}

We set the internuclear distance at $10^{9} \si{\angstrom}$ to make it comparable to a previous study by the authors using the linear and quadratic NOE-CC ansatz (l-NOECCSD and q-NOECCSD, respectively). 
\begin{table}[h!]
\renewcommand{\arraystretch}{1.5}
\begin{tabular}{ c c c c }
\hline
\hline
                                        &       l-NOECCSD           &       q-NOECCSD        &       GNOCCSD       \\
\hline
Li ($^{2}S$)                            &       -41.694641          &       -41.545784       &       -41.5461088        \\
Li$_{2}$ ($^{3}\Sigma_{g}^{+}$)            &       -83.389282          &       -83.091498       &       -83.0922176        \\
Li$_{2}$ ($^{1}\Sigma_{g}^{+}$)            &       -83.389289          &       -83.091947       &       -83.0922177        \\
$ \Delta E (^{3}\Sigma_{g}^{+})$      &        0.000001           &        -0.000070       &       $< 1 \times 10^{-7}$        \\
$ \Delta E (^{1}\Sigma_{g}^{+})$      &        0.000007           &         0.000379       &       $< 1 \times 10^{-6}$        \\
\hline
\hline
\end{tabular}
\caption{Correlation energies (in milli-Hartrees) of atomic lithium, both $M_S = 0$ spin states of the well-separated lithium dimer, and their size-inconsistency errors. GNOCCSD finds size-consistent energies for both singlet and triplet states to nano-Hartree accuracy. l-NOECCSD and q-NOECCSD data were obtained from reference \inlinecite{gunasekeraMultireferenceCoupledCluster2024}. }
\label{tab:Li2Energies}
\end{table}
In Table \ref{tab:Li2Energies}, we compare the size consistency of the NOECCSD and GNOCCSD ansätze. With the NOECCSD ansatz, it was found that the inclusion of spectator excitations led to non-additively separable cluster amplitudes and, in turn, size-inconsistent results. 
However, this problem is addressed in this work through our redundancy handling method. In this work, we have limited our excitation basis to those leading into the FOIS. The excitation basis is therefore separable (the number of non-redundant excitations in the lithium dimer is exactly twice the number of non-redundant excitations in the lithium atom), and the correlation energy is size-consistent to sub-nE$_{\text{H}}$, which is the current limit of our numerical precision and convergence thresholds.\\

\subsection{High-spin open-shell cases}
We now turn our attention to the application of GNOCC to other open-shell systems. Recently, Herrmann and Hanrath have developed the spin-adapted and spin-complete coupled cluster (SASC-CC) theory to correlate high-spin open shell systems\cite{herrmannGenerationSpinadaptedSpincomplete2020,herrmannCorrectlyScalingRigorously2022,herrmannAnalysisDifferentSets2022}. Unlike GNOCC where equations are truncated at second order in cluster amplitudes with truncation of 5- and higher body cumulants, SASC-CC equations are expanded up to quadruply nested commutators, possessing terms up to fourth order in cluster amplitudes. The SASC-CC equations are therefore more complete, and their results will serve as important benchmarks for the applicability of GNOCC to open-shell systems.\\
We present the correlation energies of various small molecules with a particular spin, $S$, and compare our results to those reported in reference \inlinecite{herrmannCorrectlyScalingRigorously2022}. For each of these calculations, the orbitals were found via a ROHF calculation and the $M_S$-averaged spin ensemble was constructed.\\
We note a caveat that while GNOCC is $M_S$ independent because a $M_S$-averaged spin ensemble is used as reference, the SASC-CC method uses a Slater determinant with $S = M_S$. Nonetheless, the energy of a spin-ensemble and that of a single spin state with a particular $M_S$ are equivalent in an exact theory, and we therefore expect the results to be comparable.\\
Before examining the results, it is helpful to note that there are two key differences between SASC-CC and GNOCC. Firstly, the definition of the cluster operator $\hat{T}$ used in both methods is different. For example, under the singles and doubles approximation, $\hat{T}$ in GNOCCSD only includes one-electron and two-electron excitations. However, $\hat{T}$ in SASC-CCSD includes some three- and four-electron excitations to achieve spin-completeness. \\
Secondly, both methods differ in the level of approximation used. While we truncate any terms involving $k$-cumulants where $k > 4$ and keep only terms up to quadratic order in cluster amplitudes, SASC-CC truncates at four-nested commutators, meaning that terms up to quartic order in cluster amplitudes are present. However, numerical examples by Herrmann and Hanrath\cite{herrmannCorrectlyScalingRigorously2022} demonstrated that even when truncating at two-nested commutators, therefore keeping terms up to quadratic order in cluster amplitudes similar to GNOCC, the energy found only differed by a few $\mu$E$_{\text{H}}$. Moreover, we do not expect the high-order cumulants to have a significant impact on the energies. Therefore, we expect that the discrepancies in results between SASC-CC and GNOCC stem from the difference in definition of the singles and doubles operators.\\
From Table \ref{tab:SASCEnergies}, we find that the correlation energies found with GNOCCSD are generally in good agreement with those found with SASC-CCSD, with most values agreeing to sub-mE$_{\text{H}}$. There are several values for which GNOCCSD and SASC-CCSD are nearly equivalent, namely $S=0$ for NH and CH$_{2}$, $S=3/2$ BeH, and $S=5/2$ BeH.\\
For the $S=0$ cases, the reference functions are closed-shell RHF states. There are no active orbitals defined in this case because all orbitals are either doubly occupied or unoccupied. The singles and doubles excitations for GNOCCSD and SASC-CCSD, which differed only due to the presence of certain spectator excitors in SASC-CCSD, are equivalent in the limit of no active orbitals. The $\mu$E$_{\text{H}}$ discrepancies in the energies calculated are due to the differing equation truncation scheme between the two methods.\\
For the case of $S=5/2$ BeH, all 5 electrons present are in the active space, and therefore there are no core orbitals. In this case, the singles and doubles excitations for GNOCCSD and SASC-CCSD are also the same, and this is reflected in the $\mu$E$_{\text{H}}$ agreement of the correlation energies.\\
For the case of $S=3/2$ BeH, 3 electrons are present in the active space, leaving 1 core orbital. Here, SASC-CCSD has an additional excitation of the form $\hat{E}_{tui}^{abt}$. The $\mu$E$_{\text{H}}$ energy difference in this case reflects the small contribution of this particular excitation.\\
These observations also justify our assumptions that the high-order cumulants and truncation of equations to quadratic order in cluster amplitudes do not affect our computed correlation energies appreciably.\\
As the number of core electrons (and therefore orbitals) increases, the number of triple excitations incorporated in the SASC-CCSD excitations becomes larger. For example, we can compare $S=1/2$ for BeH and OH, which share the same number of active space orbitals and electrons, but OH has two more core orbitals than BeH. We observe that the discrepancy between GNOCCSD and SASC-CCSD is an order of magnitude larger for OH than BeH, with the discrepancy growing to $0.2$mE$_{\text{H}}$ for OH. A similar trend is observed when comparing the $S=3/2$ cases for these two molecules. The larger discrepancy between GNOCCSD and SASC-CCSD with increasing number of electrons indicates the importance of triple excitations. The largest deviations between GNOCCSD and FCI (up to 5mE$_{\text{H}}$ for $S=0$ CH$_2$) are for the low spin states, which have more singlet coupled electron pairs. To reach chemical accuracy for all states it will be necessary to include triple excitations to parameterise three-body correlation and this will be considered in a future work.

\begin{table}[h!]
\renewcommand{\arraystretch}{1.5}
\begin{tabular}{ c c c c c }
\hline
\hline
Molecule                 &		$S=S_{z}$        &       GNOCCSD		&  SASC-CCSD$^{a}$	&		FCI$^{b}$	\\
\hline
	    		         &		1/2		         &       -0.039143	    &	-0.039178		&       -0.039797                        \\
BeH	    		         &		3/2	             &       -0.009197      &	-0.009198		&       -0.009229                        \\
    		             &	    5/2			     &       -0.015599	    &	-0.015601		&       -0.015715                        \\
\hline
		                 &		  0		         &       -0.153303	    &	  -0.153304		  &       -0.168572                        \\
NH		                 &		  1		         &       -0.131886      &	  -0.132161		  &       -0.133888                        \\
		                 &		  2		         &       -0.097770	    &	-0.098107		&       -0.099145                        \\
\hline
		                 &		1/2		         &      -0.169356	    &	  -0.169504		  &       -0.171556                        \\
OH		                 &		3/2		         &      -0.137713 	    &	  -0.137953		&       -0.139253                      \\
\hline
		                 &		  0		         &      -0.138223 	    &	  -0.138224		  &       -0.143480                        \\
CH$_{2}$		         &		  1		         &      -0.120570 	    &	  -0.120938		&       -0.122858                      \\
		                 &		  2		         &      -0.113202 	    &	  -0.114401		  &       -0.117450                        \\
\hline
\hline
\end{tabular}
\caption{Correlation energies (in Hartrees) of various high-spin open-shell systems. Geometries of all molecular systems follow those found in reference \inlinecite{herrmannCorrectlyScalingRigorously2022}. All calculations were performed with the cc-pVDZ basis. $^{a}$SASC-CCSD refers to the rigorously spin-adapted open-shell CCSD formulation of Herrmann and Hanrath. $^{a,b}$ The values were taken from reference \inlinecite{herrmannCorrectlyScalingRigorously2022}.}
\label{tab:SASCEnergies}
\end{table}

\subsection{BeH$_2$}
\begin{figure}
\includegraphics[scale=0.7]{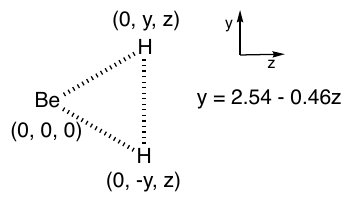}
\caption{The dissociation of BeH$_2$. Be lies on the origin, H atoms have coordinates (0, $\pm (2.54-0.46z)$, $z$).}
\label{fig:BeH2_Dissoc}
\end{figure}

\begin{figure}
\includegraphics[scale=0.55]{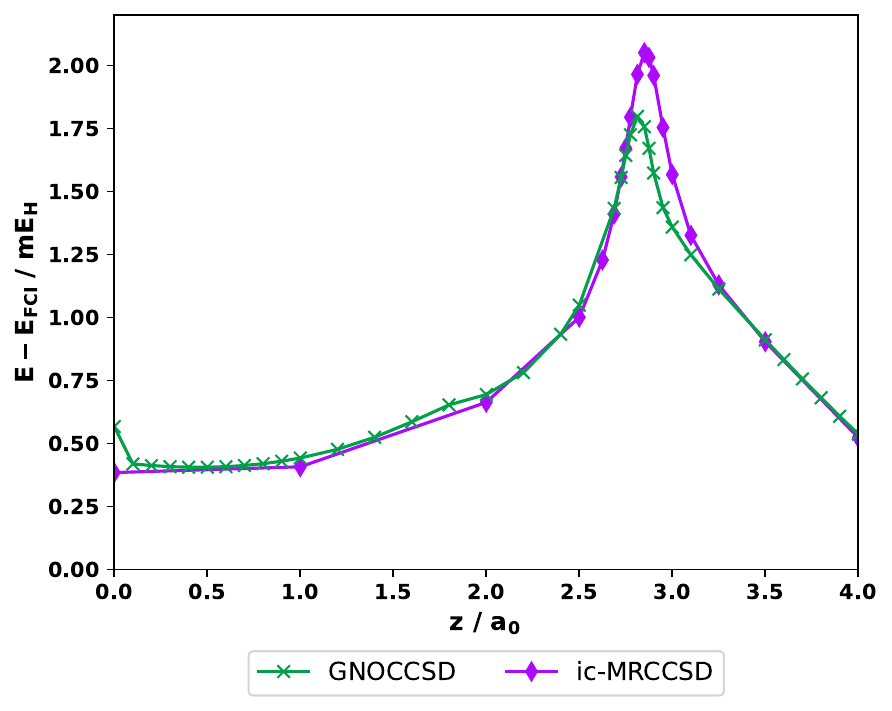}
\caption{Error of computed energies against FCI of the $^{1}A_{1}$ state of BeH$_2$ for GNOCCSD and ic-MRCCSD. The cc-pVDZ basis was used. Energies are given in milli-Hartrees, and the internuclear distances are given in atomic units. ic-MRCCSD results were taken from reference \cite{hanauerPilotApplicationsInternally2011}.}
\label{fig:BeH2_Error}
\end{figure}

The dissociation of BeH$_2$ (Figure \ref{fig:BeH2_Dissoc}) is a standard example of a process that requires a multireference treatment due to the multi-configurational character of the transition state. Therefore, this system serves as a test bed for any proposed multi-determinant CC methods. We use the BeH$_2$ dissociation model first proposed by Purvis et al.\cite{purvisiiiC2VInsertionPathway1983} where Be is placed at the origin, and each of the H atoms have coordinates ($\pm (2.54-0.46z)$, $z$), where $z$ (in Bohr) is the perpendicular distance between the Be atom and the line intersecting both hydrogen atoms. We study the dissociation of BeH$_2$ throughout the range $0 \leq z \leq 4 a_0$ using a cc-pVDZ basis. \\
Throughout the dissociation, the orbitals $1a_1$ and $2a_1$ remain doubly occupied and the active space consists of the orbitals $3a_1$ and $1b_2$. We follow the procedure of Hanauer and Köhn\cite{hanauerPilotApplicationsInternally2011} and freeze the $1a_1$ orbital in the GNOCCSD calculations to make our results comparable. \\
We investigated the error of GNOCCSD energies with respect to the FCI\cite{test} energy (Figure \ref{fig:BeH2_Error}). The errors from GNOCCSD are very similar to those of ic-MRCC. Overall, our results show a maximal absolute deviation (MAD) of 1.8 mE$_{\text{H}}$, and a non-parallelism error (NPE) of 1.3 mE$_{\text{H}}$. The NPE quantifies the consistency of deviation of the calculated energies from FCI energies, while the MAD quantifies the most significant deviation of the calculated energies from FCI energies. \\
The error peaks near the transition state ($z \approx 2.8 a_{0}$), with that of GNOCCSD $\approx$ 0.2 mE$_{\text{H}}$ smaller. As previously noted by Hanauer and Köhn\cite{hanauerPerturbativeTreatmentTriple2012}, the error stems from the lack of triple excitations in the cluster operator. The slightly smaller error in GNOCCSD can be rationalised by the difference in cluster excitations used between GNOCCSD and ic-MRCCSD. Purely active-to-active excitations are omitted in ic-MRCCSD but retained in GNOCCSD. Therefore, additional \emph{disconnected} three- and four-electron operator contributions containing spectator excitations such as $\{ t_{t}^{u} t_{ab}^{ij} \hat{E}^{t}_{u} \hat{E}^{ab}_{ij}\}$ and $\{ t_{tu}^{vw} t_{a}^{i} \hat{E}^{tu}_{vw} \hat{E}^{a}_{i}\}$ are present in GNOCCSD. In addition, there are also more independent cluster amplitudes (these correspond to purely active-to-active excitations) to parameterise these three- and four-electron excitations, affording them a better description and therefore a smaller error in the region where triple excitations are important.


\subsection{HF}
\begin{figure}
\includegraphics[scale=0.55]{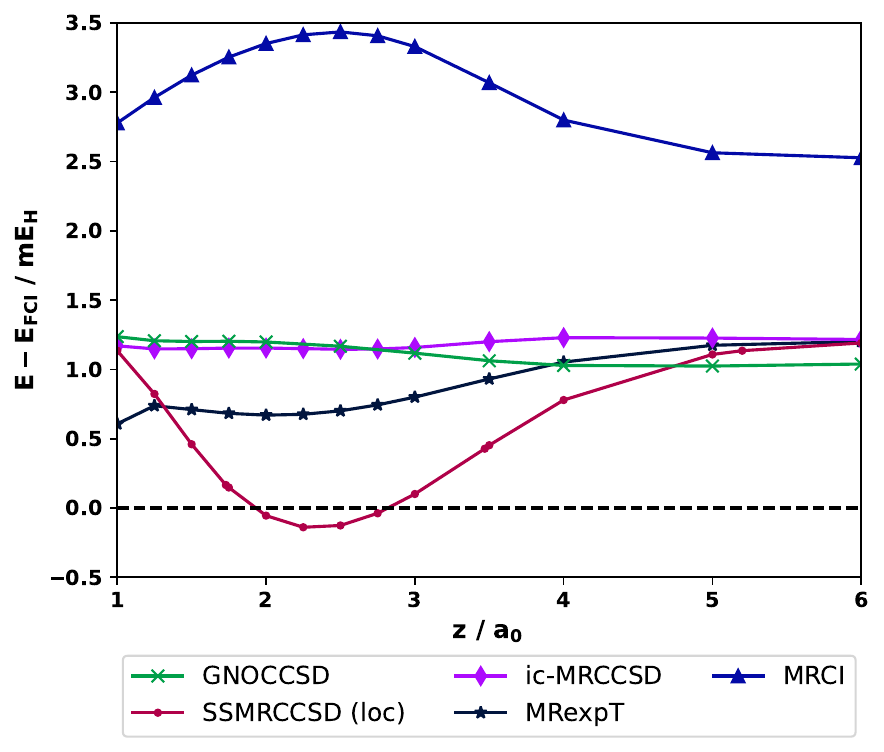}
\caption{Error of computed energies against FCI of the $^{1}A_{1}$ state of HF for various multireference methods. The DZV basis was used. Energies are given in milli-Hartrees, and the H$-$F bond length, $z$, is given in atomic units. ic-MRCCSD results were taken from reference \inlinecite{hanauerPilotApplicationsInternally2011}. MRCI, MRexpT, and FCI results were taken from reference \inlinecite{engels-putzkaMultireferenceCoupledclusterStudy2009}. SSMRCC (with localised orbitals) results were taken from reference \inlinecite{dasExternallycorrectedSizeextensiveSingleroot2006}}.
\label{fig:HF_Error}
\end{figure}

The dissociation of HF is another prototypical system for assessing multi-determinantal coupled cluster approaches. Comparative data is available for a wide range of methods using the DZV basis, following Engels-Putzka and Hanrath\cite{engels-putzkaMultireferenceCoupledclusterStudy2009}. Throughout the dissociation, the $1a_1$, $2a_1$, $1e_{1x}$, and $1e_{1y}$ orbitals remain doubly occupied, and the active space consists of the $3a_1$ and $4a_1$ orbitals.\\
We compared the error of computed energies against FCI of the $^{1}A_{1}$ state of HF for GNOCCSD, ic-MRCCSD, MRCI, SS-MRCCSD with localised orbitals and MRexpT (Figure \ref{fig:HF_Error}). For all methods, the accuracy is primarily limited by the neglect of three-body correlations. For the GNOCCSD method, the NPE is 0.2 mE$_{\text{H}}$ and the MAD is 1.2 mE$_{\text{H}}$. GNOCCSD and ic-MRCC exhibit similar levels of non-parallelity, which is significantly lower than that of the other multireference methods shown. In terms of absolute errors with respect to FCI, GNOCCSD is also shown to give similar values to ic-MRCCSD. At smaller bond lengths, methods such as MRexpT\cite{hanrathExponentialMultireferenceWavefunction2005,hanrathExponentialMultireferenceWavefunction2008,hanrathHigherExcitationsExponential2008} and SSMRCC\cite{mahapatraSizeconsistentStatespecificMultireference1999,dasFullImplementationBenchmark2010} (using localised orbitals) outperform GNOCCSD in terms of accuracy. However, as the bond length increases, GNOCCSD is shown to be slightly more accurate. This can be attributed to the use of localised orbitals, which are better suited for describing the correlation process at dissociative regimes.

\subsection{Singlet-Triplet gaps in Benzynes}

\begin{table}[h!]
\renewcommand{\arraystretch}{1.5}
\begin{tabular}{ c c c c }
\hline
\hline
            &       MkCCSD           &       ic-MRCCSD           &       GNOCCSD        \\
\hline
ortho       &       35.1            &       33.684              &       33.539        \\
meta        &       18.7            &       17.366              &       17.222        \\
para        &       4.5             &       3.586               &       3.905        \\
\hline
\hline
\end{tabular}
\caption{Singlet-Triplet gaps (in kcal/mol) of ortho-, meta-, and para-benzyne using various coupled cluster methods. Geometries of all molecular systems were taken from reference \inlinecite{evangelistaCouplingTermDerivation2007}. All calculations were performed with the cc-pVDZ basis.}
\label{tab:BenzyneEnergies}
\end{table}

Obtaining accurate singlet-triplet (ST) gaps is one important application of open-shell correlation methods. We examine the efficacy of our method in obtaining ST gaps by computing these values for ortho, meta, and para benzyne isomers. These isomers are useful test cases for multireference electronic structure methods, as they exhibit a range of diradical character, and experimental data on benzyne singlet-triplet gaps are available. We performed calculations on these isomers in a cc-pVDZ basis. The benzyne geometries were obtained from Evangelista et al.\cite{evangelistaCouplingTermDerivation2007} For ortho-benzynes and meta-benzynes, the active space comprises the $a_{1}$ and $b_{2}$ orbitals, while for para-benzynes, the active space comprises the $b_{2u}$ and $a_{g}$ orbitals. For the triplet states, the CASSCF calculations were performed with $M_S = 1$, and a spin-lowering operator was applied to find the $M_{S} = 0$ triplet which was used as the reference function. These ST gaps, computed using the same geometries and basis set, were previously reported by Evangelista et al. using MkMRCC and by Köhn et al. with ic-MRCCSD. All three methods give similar ST gaps (Table \ref{tab:BenzyneEnergies}). In particular, we note that our ST gaps are in close agreement with ic-MRCCSD, with all of our values within 0.4 kcal/mol. The results highlight that our spin-free and size-consistent coupled cluster method has accuracy comparable to existing state-of-the-art methods. 

\section{Conclusions and Outlook}
We have formulated a spin-free coupled cluster method based on a generalised normal ordered exponential ansatz. This ansatz is a natural generalisation of the traditional exponential ansatz within single-reference coupled cluster theory, where each correlation process is parameterised through an independent cluster amplitude tailored to the given reference function.
Within this framework, we can correlate arbitrary spin eigenfunctions, such as configuration state functions or CASSCF wavefunctions, allowing one to treat open-shell and multi-determinantal systems at a coupled cluster level of theory. This provides a way of performing state-specific coupled cluster on both ground and excited states that is spin-free, size-extensive, and size-consistent.\\
The GNOCC method is made spin-free by employing a spin-ensemble reference. Due to the GNO formalism, the working equations are connected and therefore size-extensive. Size-consistency is attained by the use of localised orbitals and a scheme to use linearly independent excitations leading into the first-order interacting space. This eliminates spurious excitations, which have been previously found to cause size-inconsistency. At the same time, the use of a linearly independent set of excitations resolves the problem of redundancies, which plagues many multi-reference methods. However, the redundancy handling procedure is not invariant to orbital rotations within the active space. Under the reasonable assumption that the number of active orbitals is much smaller than the number of core and virtual orbitals, the computational scaling of GNOCCSD is the same as that of CCSD.  \\
In our implementation, the working equations were truncated to second order in cluster amplitudes and to fourth order in cumulant rank. Numerical tests on a selection of open-shell and multireference systems commonly used to examine MRCC methods show that the loss of accuracy from the truncation scheme is orders of magnitude smaller than from the neglect of triple excitations. We have also numerically demonstrated size-consistency of our method through an example with the lithium dimer. Our method therefore delivers comparable accuracy to extant MRCC methods, yet retaining the appealing feature that no contractions between cluster operators occur, leading to a simpler and cleaner form of the working equations. \\
A drawback of the method lies in its unfavourable scaling with the number of active orbitals, which will limit the size of the active space that can be correlated. However, it is possible that many of these computationally expensive terms can be omitted without significant loss of accuracy.\\
The main deficiency in our current approach lies in the neglect of three-body correlation. Looking forward, we aim to include these through a perturbative triples correction. This natural generalisation of CCSD(T) to open-shell systems, will provide a unified theoretical framework for us to treat both closed-shell and open-shell systems at a consistent level of theory.


\section{Acknowledgements}
NL would like to thank Nils Herrmann for providing data regarding SASC-CC results.

\section*{References}
\bibliographystyle{aipnum4-1}
\bibliography{main}

\begin{thebibliography}{121}
\bibitem [{tes()}]{test}%
  \BibitemOpen
  \href@noop {} {}\bibinfo {note} {The FCI energy here assumes that the Be
  CASSCF 1s orbital is frozen in the correlation treatment. This is the
  approach taken in reference 39.}\BibitemShut {Stop}%
\end{thebibliography}


\newpage
\clearpage

\onecolumngrid
\appendix
\section{Justification for using spin-ensemble approach}
\label{Appendix:SpinEnsemble}
We demonstrate in this section the equivalence between a spin-ensemble formalism and the use of spin-free equations for GNOCC. Similar proofs have previously been provided.\cite{kutzelniggIrreducibleBrillouinConditions2002,shamasundarCumulantDecompositionReduced2009}\\
For a system with spin $S$, its corresponding ensemble average is given by:
\begin{equation}
\ket{\Psi} = \frac{1}{\sqrt{2S + 1}} \sum_{M=-S}^{M=S} \ket{S, M}
\end{equation}
The expectation value of an operator $\hat{O}$ with respect to a spin ensemble can be expressed as
\begin{equation}
\begin{split}
\braket{ \Psi | \hat{O} | \Psi } &= \frac{1}{2S+1} \sum_{M_1=-S}^{S} \sum_{M_2=-S}^{S}  \braket{ S, M_1 | \hat{O} | S, M_2 } \\
											   &= \frac{1}{2S+1} \sum_{M_1=-S}^{S} \sum_{M_2=-S}^{S}  \sum_{kq} c_{kq} \braket{ S, M_1 | \hat{T}_{kq} | S, M_2 } \quad\quad	\text{(Completeness of spherical tensor basis)}\\
											   &= \frac{1}{2S+1} \sum_{M_1=-S}^{S} \sum_{M_2=-S}^{S}  \sum_{kq} c_{kq} \braket{ S, M_1 | S, k, M_2, q } \braket{ S || \hat{T}_{k} || S } \quad\quad	\text{(Wigner-Eckart theorem\cite{wignerGroupTheoryIts2013,brinkAngularMomentum1994})}\\
\end{split}
\end{equation}
$\braket{ S || \hat{T} || S }$ is known as the \emph{reduced matrix element}, a term that is \emph{independent} of the spin projection of $\hat{T}$. It can therefore be taken out of the summations. For our purposes, we are only working with $M_{S}$-conserving operators (No net alpha or beta electrons created/annihilated by the operator). Consequently, $q = 0$.
Therefore,
\begin{equation}
\begin{split}
\braket{ \Psi | \hat{O} | \Psi }  &= \frac{1}{2S+1} \sum_{k} \braket{ S || \hat{T}_{k} || S } \sum_{M_1=-S}^{S} c_{k0} \braket{ S, M_1 | S, k, M_1, q }  \\
												&= \frac{1}{2S+1} \sum_{k}  \braket{ S || \hat{T}_{k} || S } c_{k0} (2S+1) \delta_{k0} \\
												&=  \braket{ S || \hat{T}_{0}|| S } c_{00}  \\
\end{split}
\end{equation}
$\hat{T}_{0}$ is a singlet tensor operator. For example, if $\hat{O} = a^{\dagger}_{p} a_{q}$, $\hat{T}_{0} = \hat{E}^{p}_{q} = a^{\dagger}_{p\alpha} a_{q\alpha} + a^{\dagger}_{p\beta} a_{q\beta}$.\\
In coupled cluster, we are concerned with evaluating equations of the type:
\begin{equation}
E = \bra{\Phi} \hat{H} \{ e^{\hat{T}} \} \ket{\Phi}
\end{equation}
\begin{equation}
0 = \bra{\Phi}  \hat{\tau}_{I}^{\dagger} \hat{H} \{ e^{\hat{T}} \}\ket{\Phi} 
\end{equation}
For the energy equation, both $\hat{H}$ and $\hat{T}$ can be spin-adapted using spin-replacement rules. For the amplitude equation, we can apply the spin-replacement rules to $\hat{H}$ and $\hat{T}$ as usual. We are unable to do the same for $\hat{\tau}_{I}^{\dagger}$ as it is an operator that is not premultiplied by an antisymmetric tensor. However, since $\hat{H}$ and $\hat{T}$ are both singlet operators after the application of spin-replacement rules, and the reference states are spin-ensemble averages, only the singlet component of  $\hat{\tau}_{I}^{\dagger}$ survives. Therefore, the amplitude equation is equivalent to
\begin{equation}
0 = \bra{\Phi}  \hat{E}_{I}^{\dagger} \hat{H} \{ e^{\hat{T}} \}\ket{\Phi} 
\end{equation}
where $\hat{E}_{I}$ is an arbitrary singlet excitation operator.

\clearpage
\section{Derivation of working equations}
\label{Appendix:EquationsDerivation}
We derive the energy and residual equations here. The equations\cite{mahapatraStateSpecificMultiReferenceCoupled1998} were first given by Mukherjee and later expanded upon in Mukherjee\cite{sinhaGeneralizedAntisymmetricOrdered2013}. We begin from the Schrödinger equation and the wavefunction ansatz
\begin{equation}
\Psi = \{ e^{\hat{T}} \} \Phi
\end{equation}
From the Schrödinger equation,
\begin{equation}
\begin{split}
\hat{H} \ket{\Psi} 	&=	E \ket{\Psi} \\
\hat{H} \{ e^{\hat{T}} \} \ket{\Phi}	&=	E \{ e^{\hat{T}} \} \ket{\Phi} \\
\{ e^{\hat{T}} (\hat{H} \{ e^{\hat{T}} \} )_{c} \} \ket{\Phi}	&=	E \{ e^{\hat{T}} \} \ket{\Phi} \\
\end{split}
\label{eqn:ContractedSE}
\end{equation}
The last line is found by applying the well-known result\cite{lindgrenCoupledclusterApproachManybody1978,mukherjeeCorrelationProblemOpenshell1975} that 
\begin{equation}
    \hat{H} \{ e^{\hat{T}} \} = \{ e^{\hat{T}} ( \hat{H} e^{\hat{T}} )_{c}     \}
\end{equation}  
For the energy equation, we project the equation onto our reference state $\bra{\Phi}$:
\begin{equation}
\begin{split}
\bra{\Phi} \{ e^{\hat{T}} (\hat{H} \{ e^{\hat{T}} \} )_{c} \} \ket{\Phi}		&=	E \bra{\Phi} \{ e^{\hat{T}} \} \ket{\Phi} \\
																										&=	E	\quad\quad \text{(Wick's theorem)}
\end{split}
\end{equation}
Within $\bra{\Phi} \{ e^{\hat{T}} (\hat{H} \{ e^{\hat{T}} \} )_{c} \} \ket{\Phi}$, terms such as $\bra{\Phi} \{ \hat{T}^{n} (\hat{H} \{ e^{\hat{T}} \} )_{c} \} \ket{\Phi}$ are necessarily zero because there are uncontracted terms in the normal-ordered operator. However, it is possible for $(\hat{H} \{ e^{\hat{T}} \})_{c}$ to contain \emph{fully contracted} terms which will contribute to a non-zero expectation value. Therefore,
\begin{equation}
E = \bra{\Phi} \{ e^{\hat{T}} (\hat{H} \{ e^{\hat{T}} \} )_{c} \} \ket{\Phi}	 = \bra{\Phi} \{ (\hat{H} \{ e^{\hat{T}} \} )_{c} \} \ket{\Phi}
\end{equation}
To arrive at the residual equation, we project the equation onto excited references $\bra{\Phi} \hat{\tau}^{\dagger}_{\mu}$ to find
\begin{equation}
\begin{split}
\bra{\Phi} \hat{\tau}^{\dagger}_{\mu}\{ e^{\hat{T}} (\hat{H} \{ e^{\hat{T}} \} )_{c} \} \ket{\Phi}		&=	E \bra{\Phi} \hat{\tau}^{\dagger}_{\mu} \{ e^{\hat{T}} \} \ket{\Phi} \\\
						&=	\bra{\Phi} \{ (\hat{H} \{ e^{\hat{T}} \} )_{c} \} \ket{\Phi} \bra{\Phi} \hat{\tau}^{\dagger}_{\mu} \{ e^{\hat{T}} \} \ket{\Phi} \\
						&=	\bra{\Phi} \{ (\hat{H} \{ e^{\hat{T}} \} )_{c} \} \ket{\Phi} \bra{\Phi} \{ e^{\hat{T}} (\hat{\tau}^{\dagger}_{\mu} \{ e^{\hat{T}} \} )_{Cc} \} \ket{\Phi} \\
						&=	\bra{\Phi} \{ (\hat{H} \{ e^{\hat{T}} \} )_{c} \} \ket{\Phi} \bra{\Phi} \{ (\hat{\tau}^{\dagger}_{\mu} \{ e^{\hat{T}} \} )_{c} \} \ket{\Phi} \\
\end{split}
\label{eqn:NaiveAmplitudeEqn}
\end{equation}
The RHS can be derived using similar arguments to those applied to the derivation of the energy equation. For the LHS, $\hat{\tau}^{\dagger}_{\mu}$ has to contract with $\{ e^{\hat{T}} (\hat{H} \{ e^{\hat{T}} \} )_{c} \} $ as only the fully contracted terms survive. There are only 3 possibilities:
\begin{enumerate}
\item $\hat{\tau}^{\dagger}_{\mu}$ contracts only with $e^{\hat{T}} $. $e^{\hat{T}} $ here refers \emph{only} to the exponential operator to the left of $(\hat{H} \{ e^{\hat{T}} \} )_{c}$
\item $\hat{\tau}^{\dagger}_{\mu}$ contracts only with $(\hat{H} \{ e^{\hat{T}} \} )_{c}$
\item $\hat{\tau}^{\dagger}_{\mu}$ contracts with both $e^{\hat{T}} $ and $(\hat{H} \{ e^{\hat{T}} \} )_{c}$
\end{enumerate}
Examining case 1, we find that for the expression to be non-zero when $\hat{\tau}^{\dagger}_{\mu}$ contracts only with $e^{\hat{T}} $, the expressions can only contain full-contracted parts of $(\hat{H} \{ e^{\hat{T}} \} )_{c}$. The fully contracted part can also be written as $\bra{\Phi} \{ (\hat{H} \{ e^{\hat{T}} \} )_{c} \} \ket{\Phi} = E$
\begin{equation}
\bra{\Phi} \hat{\tau}^{\dagger}_{\mu} \{ e^{\hat{T}} (\hat{H} \{ e^{\hat{T}} \} )_{c} \} \ket{\Phi} \rightarrow \bra{\Phi} \hat{\tau}^{\dagger}_{\mu} \{ e^{\hat{T}} \} \ket{\Phi} \bra{\Phi}  \{ (\hat{H} \{ e^{\hat{T}} \} )_{c} \} \ket{\Phi} = \bra{\Phi} \{ (\hat{\tau}^{\dagger}_{\mu} \{ e^{\hat{T}} \} )_{c} \} \ket{\Phi}  \bra{\Phi}  \{ (\hat{H} \{ e^{\hat{T}} \} )_{c} \} \ket{\Phi} 
\end{equation}
We find that Case 1 is simply RHS of equation \ref{eqn:NaiveAmplitudeEqn}. Subtracting the RHS on both sides of equation \ref{NaiveAmplitudeEqn}, we get
\begin{equation}
\text{Case 2} + \text{Case 3} = \bra{\Phi} \{ ( \hat{\tau}^{\dagger}_{\mu} \hat{H} \{ e^{\hat{T}} \} )_{c} \} \ket{\Phi}	 = 0
\label{AmplitudeEqn}
\end{equation}
This is our working amplitude equation.

\clearpage

\section{Overlap matrices for redundancy handling}
We detail the expressions to compute spin-free overlap matrices for various excitation classes.
\label{Appendix:RedundancyOverlap}
\subsection{Overlaps for redundancy handling -- Single Excitations}
\subsubsection{$\mathbb{C} \rightarrow \mathbb{A}$}
\begin{equation}
\begin{split}
S^{i,v}_{u,j} 	&= \braket{\Phi | \tilde{E}^{u}_{i}{}^{\dagger}  \tilde{E}^{v}_{j} | \Phi} \\
					&= \delta^{i}_{j} \Theta^{v}_{u}
\end{split}
\end{equation}

\subsubsection{$\mathbb{A} \rightarrow \mathbb{V}$}
\begin{equation}
\begin{split}
S^{t,b}_{a,u} 	&= \braket{\Phi | \tilde{E}^{a}_{t}{}^{\dagger}  \tilde{E}^{b}_{u} | \Phi} \\
					&= \delta^{b}_{a} \Gamma^{t}_{u}
\end{split}
\end{equation}

\subsubsection{$\mathbb{A} \rightarrow \mathbb{A}$}
\begin{equation}
\begin{split}
S^{u,x}_{v,w} 	&= \braket{\Phi | \tilde{E}^{v}_{u}{}^{\dagger}  \tilde{E}^{x}_{w} | \Phi} \\
						&= \frac{1}{2} \Gamma^{u}_{w} \Theta^{x}_{v} + \Lambda^{ux}_{vw}
\end{split}
\end{equation}

\subsection{Overlaps for redundancy handling -- Double Excitations}
\subsubsection{$\mathbb{CA} \rightarrow \mathbb{AV}$}
\begin{equation}
\begin{split}
S^{iu,xb}_{va,jw} 	&=	\braket{\Phi | \tilde{E}^{va}_{iu}{}^{\dagger}  \tilde{E}^{xb}_{jw} | \Phi} \\
							&=	\delta^{i}_{j} \delta^{b}_{a} ( \Gamma^{u}_{w} \Theta^{x}_{v} - \Lambda^{ux}_{wv})
\end{split}
\end{equation}

\subsubsection{$\mathbb{CA} \rightarrow \mathbb{VA}$}
\begin{equation}
\begin{split}
S^{iu,bx}_{av,jw} 	&=	\braket{\Phi | \tilde{E}^{av}_{iu}{}^{\dagger}  \tilde{E}^{bx}_{jw} | \Phi} \\
							&=	\delta^{i}_{j} \delta^{b}_{a} ( \Gamma^{u}_{w} \Theta^{x}_{v} + 2 \Lambda^{ux}_{vw})
\end{split}
\end{equation}

\subsubsection{$\mathbb{CA} \rightarrow \mathbb{VV}$}
\begin{equation}
\begin{split}
S^{iu,cd}_{ab,jv} 	&=	\braket{\Phi | \tilde{E}^{ab}_{iu}{}^{\dagger}  \tilde{E}^{cd}_{jv} | \Phi} \\
							&=	\delta^{i}_{j} \Gamma^{u}_{v} (2 \delta^{d}_{b} \delta^{c}_{a} - \delta^{d}_{a} \delta^{c}_{b})
\end{split}
\end{equation}

\subsubsection{$\mathbb{CC} \rightarrow \mathbb{AV}$}
\begin{equation}
\begin{split}
S^{ij,vb}_{ua,kl} 	&=	\braket{\Phi | \tilde{E}^{ua}_{ij}{}^{\dagger}  \tilde{E}^{vb}_{kl} | \Phi} \\
							&=	\delta^{b}_{a} \Theta^{v}_{u} (2\delta^{i}_{k} \delta^{j}_{l} - \delta^{i}_{l} \delta^{j}_{k})
\end{split}
\end{equation}

\subsubsection{$\mathbb{CC} \rightarrow \mathbb{AA}$}
\begin{equation}
\begin{split}
S^{ij,wx}_{uv,kl} 	&=	\braket{\Phi | \tilde{E}^{uv}_{ij}{}^{\dagger}  \tilde{E}^{wx}_{kl} | \Phi} \\
							&=	(\Theta^{w}_{u} \Theta^{x}_{v} - \frac{1}{2} \Theta^{w}_{v} \Theta^{x}_{u} + \Lambda^{wx}_{uv}) \delta^{i}_{k} \delta^{j}_{l} + (\Theta^{w}_{v} \Theta^{x}_{u} - \frac{1}{2} \Theta^{w}_{u} \Theta^{x}_{v} + \Lambda^{wx}_{vu}) \delta^{i}_{l} \delta^{j}_{k}
\end{split}
\end{equation}

\subsubsection{$\mathbb{CA} \rightarrow \mathbb{AA}$}
\begin{equation}
\begin{split}
S^{iu,vw}_{yz,jx} 		&=	\braket{\Phi | \tilde{E}^{vw}_{iu}{}^{\dagger}  \tilde{E}^{yz}_{jx} | \Phi} \\
								&=	-\Lambda^{uyz}_{wvx} - \frac{1}{2} \Theta^{y}_{w} \Lambda^{uz}_{vx} - \frac{1}{2} \Theta^{z}_{w} \Lambda^{uy}_{xv} - \frac{1}{2} \Theta^{z}_{v} \Lambda^{uy}_{wx} + \Theta^{y}_{v} \Lambda^{uz}_{wx} + \frac{1}{2} \Gamma^{u}_{x} (\Theta^{z}_{w} \Theta^{y}_{v} - \frac{1}{2} \Theta^{z}_{v} \Theta^{y}_{w} + \Lambda^{yz}_{vw} )
\end{split}
\end{equation}

\subsubsection{$\mathbb{AA} \rightarrow \mathbb{AV}$}
\begin{equation}
\begin{split}
S^{tu,zb}_{va,xy} 		&=	\braket{\Phi | \tilde{E}^{va}_{tu}{}^{\dagger}  \tilde{E}^{zb}_{xy} | \Phi} \\
								&=	[\frac{1}{2} \Theta^{z}_{v} (\Gamma^{t}_{x} \Gamma^{u}_{y} - \frac{1}{2} \Gamma^{t}_{y} \Gamma^{u}_{x} + \Lambda^{tu}_{xy}) - \frac{1}{2} \Gamma^{t}_{x} \Lambda^{uz}_{yv} - \frac{1}{2} \Gamma^{t}_{y} \Lambda^{uz}_{vx} + \Gamma^{u}_{y} \Lambda^{tz}_{vx} - \frac{1}{2} \Gamma^{u}_{x} \Lambda^{tz}_{vy} + \Lambda^{tuz}_{vyx} ] \delta^{b}_{a}
\end{split}
\end{equation}

\subsubsection{$\mathbb{AA} \rightarrow \mathbb{VV}$}
\begin{equation}
\begin{split}
S^{tu,cd}_{ab,vw} 		&=	\braket{\Phi | \tilde{E}^{ab}_{tu}{}^{\dagger}  \tilde{E}^{cd}_{vw} | \Phi} \\
								&=	(\Gamma^{t}_{v} \Gamma^{u}_{w} -  \frac{1}{2} \Gamma^{t}_{w} \Gamma^{u}_{v} + \Lambda^{tu}_{vw} )\delta^{d}_{b} \delta^{c}_{a} + (\Gamma^{u}_{v} \Gamma^{t}_{w} -  \frac{1}{2} \Gamma^{u}_{w} \Gamma^{t}_{v} + \Lambda^{ut}_{vw} )\delta^{c}_{b} \delta^{d}_{a} 
\end{split}
\end{equation}

\subsubsection{$\mathbb{AA} \rightarrow \mathbb{AA}$}
\begin{equation}
\begin{split}
S^{pr,tv}_{qs,uw} 		&=	\braket{\Phi | \tilde{E}^{qs}_{pr}{}^{\dagger}  \tilde{E}^{tv}_{uw} | \Phi} \\
								&=	\Lambda^{prtv}_{qsuw} + \frac{1}{2} \Theta^{v}_{s} \Lambda^{prt}_{qwu} + \frac{1}{2} \Theta^{v}_{q} \Lambda^{prt}_{wsu} + \frac{1}{2} \Theta^{t}_{s} \Lambda^{prv}_{quw} + \frac{1}{2} \Theta^{t}_{q} \Lambda^{prv}_{usw} - \frac{1}{2} \Gamma^{r}_{u} \Lambda^{ptv}_{qsw}  - \frac{1}{2} \Gamma^{r}_{w} \Lambda^{ptv}_{qus}  - \frac{1}{2} \Gamma^{p}_{u} \Lambda^{rtv}_{sqw}  - \frac{1}{2} \Gamma^{p}_{w} \Lambda^{rtv}_{suq} \\
								&\quad + \frac{1}{4} \Theta^{t}_{q} \Theta^{v}_{s} (\Gamma^{p}_{u} \Gamma^{r}_{w} - \frac{1}{2}\Gamma^{p}_{w} \Gamma^{r}_{u} + \Lambda^{pr}_{uw}) + \frac{1}{4} \Theta^{t}_{s} \Theta^{v}_{q} (\Gamma^{p}_{w} \Gamma^{r}_{u} - \frac{1}{2}\Gamma^{p}_{u} \Gamma^{r}_{w} + \Lambda^{pr}_{wu}) + \frac{1}{4} \Gamma^{p}_{u} \Gamma^{r}_{w} \Lambda^{tv}_{qs} + \frac{1}{4} \Gamma^{p}_{w} \Gamma^{r}_{u} \Lambda^{tv}_{sq} \\
								&\quad + \frac{1}{3} \Lambda^{tv}_{qs}\Lambda^{pr}_{uw} + \frac{1}{3} \Lambda^{tv}_{sq}\Lambda^{pr}_{wu} + \frac{1}{6} \Lambda^{tv}_{qs}\Lambda^{pr}_{wu} + \frac{1}{6} \Lambda^{tv}_{sq}\Lambda^{pr}_{uw} + \frac{1}{2} \Theta^{v}_{s} \Gamma^{r}_{w} \Lambda^{pt}_{qu} - \frac{1}{4} \Theta^{v}_{s} \Gamma^{r}_{u} \Lambda^{pt}_{qw} -\frac{1}{4} \Theta^{v}_{q} \Gamma^{r}_{w} \Lambda^{pt}_{su} -\frac{1}{4} \Theta^{v}_{q} \Gamma^{r}_{u} \Lambda^{pt}_{ws} \\
								&\quad -\frac{1}{4} \Theta^{t}_{s} \Gamma^{r}_{w} \Lambda^{pv}_{qu} + \frac{1}{2} \Theta^{t}_{s} \Gamma^{r}_{u} \Lambda^{pv}_{qw}  - \frac{1}{4} \Theta^{t}_{q} \Gamma^{r}_{w} \Lambda^{pv}_{us} - \frac{1}{4} \Theta^{t}_{q} \Gamma^{r}_{u} \Lambda^{pv}_{sw} 
- \frac{1}{4} \Theta^{v}_{s} \Gamma^{p}_{w} \Lambda^{rt}_{qu} - \frac{1}{4} \Theta^{v}_{s} \Gamma^{p}_{u} \Lambda^{rt}_{wq} + \frac{1}{2} \Theta^{v}_{q} \Gamma^{p}_{w} \Lambda^{rt}_{su} - \frac{1}{4} \Theta^{v}_{q} \Gamma^{p}_{u} \Lambda^{rt}_{sw} \\	
								&\quad - \frac{1}{4} \Theta^{t}_{s} \Gamma^{p}_{w} \Lambda^{rv}_{uq} - \frac{1}{4} \Theta^{t}_{s} \Gamma^{p}_{u} \Lambda^{rv}_{qw} - \frac{1}{4} \Theta^{t}_{q} \Gamma^{p}_{w} \Lambda^{rv}_{su} + \frac{1}{2} \Theta^{t}_{q} \Gamma^{p}_{u} \Lambda^{rv}_{sw} 
- \frac{1}{2} \Lambda^{pr}_{ws} \Lambda^{tv}_{uq} - \frac{1}{2} \Lambda^{pr}_{us} \Lambda^{tv}_{qw} - \frac{1}{2} \Lambda^{pr}_{qw} \Lambda^{tv}_{us} - \frac{1}{2} \Lambda^{pr}_{qu} \Lambda^{tv}_{sw}\\
								&\quad - \frac{1}{2} \Lambda^{tr}_{qs} \Lambda^{pv}_{uw} - \frac{1}{2} \Lambda^{tr}_{uw} \Lambda^{pv}_{qs}  - \frac{1}{2} \Lambda^{vr}_{qs} \Lambda^{tp}_{uw} - \frac{1}{2} \Lambda^{pt}_{qs} \Lambda^{rv}_{uw} + \Lambda^{rv}_{sw} \Lambda^{pt}_{qu} + \Lambda^{rt}_{su} \Lambda^{pv}_{qw} - \frac{1}{2} \Lambda^{rv}_{su} \Lambda^{pt}_{qw} - \frac{1}{2} \Lambda^{rv}_{qw} \Lambda^{pt}_{su} - \frac{1}{2} \Lambda^{pv}_{qu} \Lambda^{rt}_{sw} - \frac{1}{2} \Lambda^{rt}_{qu} \Lambda^{pv}_{sw} \\
								&\quad + \frac{1}{3} \Lambda^{rv}_{qu} \Lambda^{pt}_{sw} + \frac{1}{3} \Lambda^{rv}_{uq} \Lambda^{pt}_{ws} + \frac{1}{6} \Lambda^{rv}_{uq} \Lambda^{pt}_{sw} + \frac{1}{6} \Lambda^{rv}_{qu} \Lambda^{pt}_{ws} + \frac{1}{3} \Lambda^{rt}_{qw} \Lambda^{pv}_{su} + \frac{1}{3} \Lambda^{rt}_{wq} \Lambda^{pv}_{us} + \frac{1}{6} \Lambda^{rt}_{wq} \Lambda^{pv}_{su} + \frac{1}{6} \Lambda^{rt}_{qw} \Lambda^{pv}_{us}
\end{split}
\end{equation}

\subsection{Overlaps for redundancy handling -- Mixed Excitations}
\subsubsection{$\mathbb{A} \rightarrow \mathbb{V}/ \mathbb{AA} \rightarrow \mathbb{AV}$}
\begin{equation}
\begin{split}
S^{u,xb}_{a,vw} 	&=	\braket{\Phi | \tilde{E}^{a}_{u}{}^{\dagger}  \tilde{E}^{xb}_{vw} | \Phi} \\
							&=	\delta^{b}_{a} \Lambda^{ux}_{wv}
\end{split}
\end{equation}

\subsubsection{$\mathbb{C} \rightarrow \mathbb{A}/ \mathbb{CA} \rightarrow \mathbb{AA}$}
\begin{equation}
\begin{split}
S^{i,wx}_{u,jv} 	&=	\braket{\Phi | \tilde{E}^{u}_{i}{}^{\dagger}  \tilde{E}^{wx}_{jv} | \Phi} \\
						&=	-\delta^{i}_{j} \Lambda^{wx}_{uv}
\end{split}
\end{equation}

\subsubsection{$\mathbb{A} \rightarrow \mathbb{A}/ \mathbb{AA} \rightarrow \mathbb{AA}$}
\begin{equation}
\begin{split}
S^{t,yz}_{u,wx} 	&=	\braket{\Phi | \tilde{E}^{u}_{t}{}^{\dagger}  \tilde{E}^{yz}_{wx} | \Phi} \\
							&=	\Lambda^{tyz}_{uwx} - \frac{1}{2} \Gamma^{t}_{w} \Lambda^{yz}_{ux} - \frac{1}{2} \Gamma^{t}_{x} \Lambda^{yz}_{wu} + \frac{1}{2} \Theta^{y}_{u} \Lambda^{tz}_{wx} + \frac{1}{2} \Theta^{z}_{u} \Lambda^{ty}_{xw} 
\end{split} 
\end{equation}

\subsubsection{$\mathbb{CA} \rightarrow \mathbb{AV}/ \mathbb{CA} \rightarrow \mathbb{VA}$}
\begin{equation}
\begin{split}
S^{iu,by}_{wa,jx}   &= \braket{\Phi | \tilde{E}^{wa}_{iu}{}^{\dagger}  \tilde{E}^{by}_{jx} | \Phi} \\
                    &= -\delta^{i}_{j} \delta^{b}_{a} ( \frac{1}{2} \Gamma^{u}_{x} \Theta^{y}_{w} + \Lambda^{uy}_{wx})
\end{split}
\end{equation}
\end{document}